\documentstyle[eqsecnum,preprint,aps,psfig]{revtex}

\def\beq{\begin{equation}}
\def\eeq{\end{equation}}
\def\bea{\begin{eqnarray}}
\def\eea{\end{eqnarray}}
\def\nnu{\nonumber}
\def\tst{\textstyle}
\def\noi{\noindent}
\def\lbl{\label}
\def\by{\over}

\def\rno#1{Ref.~\cite{#1}}

\def\eno#1{Eq.~(\ref{#1})}

\def\al{\alpha}

\def\gam{\gamma}
\def\dta{\delta}
\def\eps{\epsilon}
\def\tta{\theta}
\def\kap{\kappa}
\def\lam{\lambda}

\def\om{\omega}

\def\Dta{\Delta}
\def\Gam{\Gamma}

\def\Om{\Omega}
\def\Lam{\Lambda}

\def\ptl{\partial}
\def\hf{{1\over2}}
\def\tshf{\tst\hf}
\def\tofro{\leftrightarrow}

\def\apx{\approx}

\def\lp{\left(}
\def\rp{\right)}

\def\ham{{\cal H}}
\def\ket#1{|#1\rangle}

\def\tran#1#2{\langle#1|#2\rangle}
\def\avg#1{\langle#1\rangle}
\def\mel#1#2#3{\langle#1|#2|#3\rangle}

\def\bA{{\bf A}}
\def\bB{{\bf B}}
\def\bH{{\bf H}}
\def\bJ{{\bf J}}

\def\xhat{{\bf{\hat x}}}
\def\yhat{{\bf{\hat y}}}
\def\zhat{{\bf{\hat z}}}
\def\nhat{{\bf{\hat n}}}
\def\Jhat{{\bf{\hat J}}}

\def\citn#1#2#3#4#5{#1, #2 {\bf#3}, #4 (#5)} 
\def\jap{J.\ Appl.\ Phys.\ }
\def\jmp{J.\ Math.\ Phys.\ }
\def\npb{Nuc.\ Phys.\ B}
\def\prb{Phys.\ Rev.\ B}
\def\prd{Phys.\ Rev.\ D}
\def\prl{Phys.\ Rev.\ Lett.\ }

\def\Fe8{Fe$_8$}
\def\tbt{{3\over 2}}
\def\comp{{\rm c.c.}}
\def\hsc{{\ham_{\rm sc}}}
\def\Mn12{Mn$_{12}$}
\def\baj{\bar J}
\def\omzm{\om_{0-}}
\def\omzp{\om_{0+}}
\def\ompm{\om_{0\pm}}
\def\Dll{\Dta({\ell',\ell''})}

\def\sla{\sqrt{\lam}}
\def\Ecl{E_{\rm cl}}
\def\Scl{S_{\rm cl}}
\def\ntau{\nhat(\tau)}
\def\kpm{\ket{\psi_{\pm}}}
\def\kp#1{\ket{\psi_{#1}}}
\def\are{{\cal A}}
\def\ant{\are[\ntau]}
\def\crv{{\cal C}}
\def\dedn{{\ptl \Ecl \by \ptl \nhat}}
\def\dndt{{d\nhat \by d\tau}}  
\def\rta{\sqrt{1-u_0^2}}

\def\rtl{\sqrt\lam}
\def\rtlb{\sqrt{1 - \lam}}
\def\rtd{\sqrt{(1-u_0^2)(1-\lam)}}

\begin{document}
\draft

\title{Spin Tunneling in Molecular Magnets}

\author{Anupam Garg}
\address{Department of Physics and Astronomy, Northwestern University,
Evanston, Illinois 60208}

\date{\today}

\maketitle

\begin{abstract}
We study spin tunneling in magnetic molecules, with special reference
to \Fe8. The article aims to give a pedagogical discussion of what
is meant by the tunneling of a spin, and how tunneling amplitudes or
energy level splittings may be calculated using path integral and
discrete phase integral methods. In the case of \Fe8, an issue of great
interest is the oscillatory tunnel splittings as a function of applied
magnetic field that have recently been observed. These oscillations
are due to the occurrence of diabolical points in the magnetic field
space. It is shown how this effect comes about in both the path-integral
and the discrete phase integral methods. In the former it arises due to
the presence of a Berry-like phase in the action, which gives rise to an
interference between tunneling trajectories. In the latter, it arises
due to the presence of further neighbor terms in the recursion relation
for the energy eigenfunction. These terms give rise to turning points
which have no analog in the one-dimensional continuum quasiclassical
method. Explicit results are obtained for the location of the
diabolical points in \Fe8.
\end{abstract}

\widetext

\section{Introduction}
\label{Intro}

Tunneling is a basic way in which the difference between quantum and
classical mechanics manifests itself, and even though the simplest examples
of tunneling were studied right after the birth of quantum mechanics, there
are many other aspects of tunneling that are still the subject of active
research today. It is generally quite difficult to calculate tunnel
splittings in many particle systems, and even for one particle, the step
from one to two spatial dimensions presents significant challenges and
surprises \cite{sc98,mw86}.

In this article, I shall study the tunneling of a single spin degree
of freedom. This is yet another instance where the statement of the problem
is very simple, yet the earliest study of which I am aware dates to 1978%
\cite{ks78}, a half-century after the founding of modern quantum theory.

From the point of view of this volume, the greatest interest in spin tunneling
lies in the possibility of observing meso or macroscopic quantum phenomena
(MQP) in magnetic particles and related systems, as first proposed in the
late 1980's \cite{cg88,bc90}. The range of activity over the next half-decade
is well represented in a workshop proceedings \cite{gb95}. After preliminary
investigation, small magnetic particles appear to be attractive candidates for
MQP, since for diameters $\sim 50$ \AA, and typical anisotropy constants, the
tunneling rates appear to be moderately large. However, in contrast to the
situation that prevails in the SQUID systems (for which, see the article by
Han in this volume), it turns out that the {\it classical\/} dynamics of the
net magnetic moment of a small particle is not fully understood theoretically,
especially with regard to dissipation. On the experimental side, the
characterization and reproducible fabrication of very small particles is still
rather difficult, and at present even the measurement and modeling of the
classical dynamics is not a fully solved problem. Wernsdorfer's article in this
volume gives a good sense of the issues involved. It is this author's opinion
(which is not necessarily shared widely), that unless the classical dynamics
is fully understood, theoretical analyses are not likely to be pertinent,
and the interpretation of experiments will be uncertain.

In the last few years, however, a much more fruitful avenue for the spin
tunneling has opened up in the area of large magnetic molecules. Several
dozen molecules are presently under study, but the two that have yielded the
most fruitful results are \Mn12 and \Fe8. In this author's view, which is
again not necessarily held by the majority of workers in the field,
the hysteresis phenomena seen in \Mn12 to date%
\cite{ns94,fs96,tlbd96} (c.f. review by
Friedman in this volume) are not due to spin
tunneling alone, but reflect a much richer and more complex many-body
effect originating in the spin-phonon interaction \cite{vhsr94,agprl98}. In the
case of \Fe8, however, there is unambiguous evidence of spin tunneling in
the form of an oscillatory magnetic field dependence of the tunnel splitting,
as seen by Wernsdorfer and Sessoli \cite{ws99}. This dependence is highly
systematic, and not easy to obtain by accident, so as an experimental
signature it is very robust. Theoretically, these oscillations represent
a very interesting difference between spin and massive particles. This
difference can be attributed almost tautologically to the difference in the
commutation relations. A much more visual representation of the difference
is provided by the difference in path integrals between these systems. The
spin path integral contains a kinetic term which has the properties of a
Berry phase, which can give rise to interference between different spin 
trajectories \cite{ldg92,vdh92}. The oscillations are a result of such
an interference effect, and were in fact predicted theoretically without
knowing of the relevance of the work to \Fe8 \cite{agepl93}. Very briefly,
here is how the effect arises. The classical state of a spin is defined
by giving its orientation $\nhat$, and the paths lie on the surface of
a unit sphere. The sum over paths is dominated by least action paths,
or {\it instantons\/}.
For certain field orientations, one finds that there are two such least
action paths that wind around $\bH$ in opposite directions (Fig. 1). 
The real part $S_R$ of the Euclidean action is equal for both paths, but the
imaginary parts differ, giving rise to a relative phase equal to the Berry phase
for the closed loop formed by the two paths. This Berry-phase is proportional
to the area $\Om$ of the loop. Thus, the splitting $\Dta$ is given by
\beq
\Dta \propto \exp(-S_R) \cos\Phi,
\eeq
where $\Phi = J\Om/2$, with $J$ being the magnitude of the spin. As the field
is increased, the minima between which the instantons run move toward each
other, and the area $\Om$ shrinks. Whenever $\Phi$ passes through an odd
multiple of $\pi/2$, $\Dta$ vanishes. In Fig. 1, we also show the result of
a direct numerical computation of $\Dta$ as a function of $H$ for the model
Hamiltonian used in Ref.~\cite{agepl93}, showing that the effect is real.
\begin{figure}
\centerline{\psfig{figure=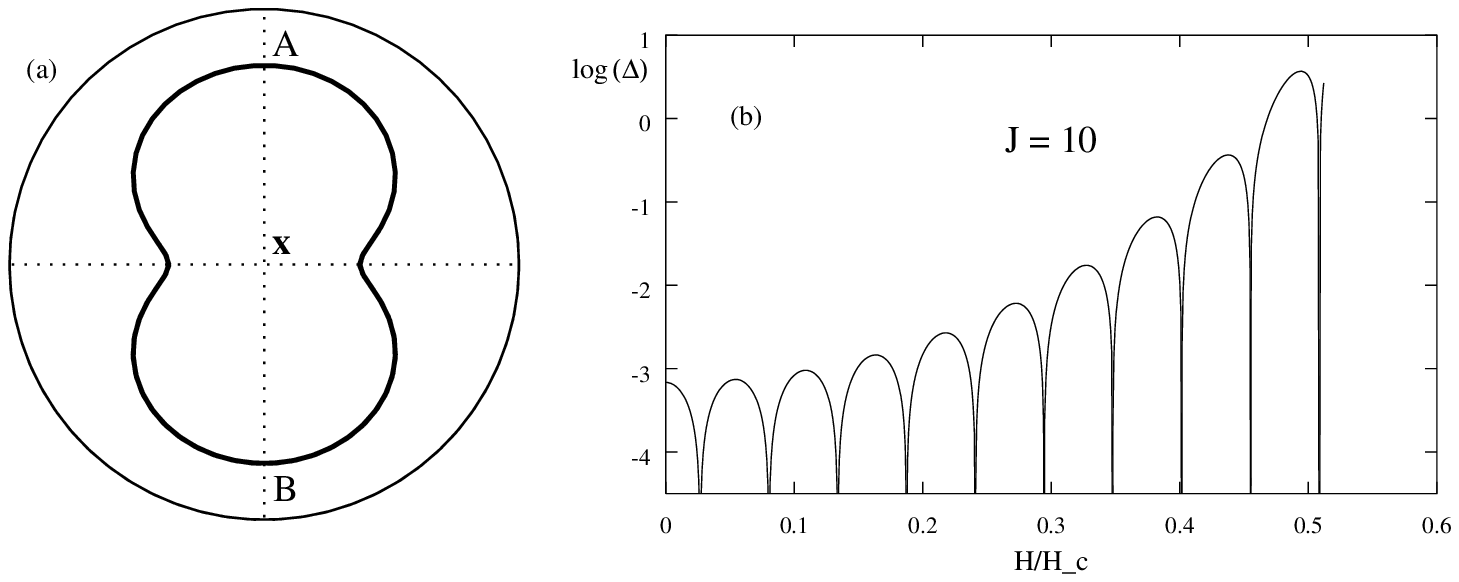}}
\caption{Interfering instanton trajectories, and numerically computed
$-10 \tofro +10$ tunnel splitting for a model for \Fe8 with $\bH\|\xhat$.}
\end{figure}

However, a proper
discussion of spin path integrals requires the development of a fair
amount of calculational machinery, which is not part of the standard
repertoire of most physicists despite having been around for almost 25
years \cite{jrk78,jrk79}. Indeed, in discussions with non-experts this
author has found that several more elementary issues need to be clarified
first. Further, path integrals are just one way to calculate tunnel
splittings. In analogy with massive particles, there is also a
phase integral or WKB method \cite{dm67,sg75,pb93,vhs86,vhw88,%
vhs95} that can be applied to spin. For some calculational purposes, this
is in fact superior to the path integral approach. Unfortunately, this
method is also not standard textbook material, and while it involves far
more elementary mathematics than the path integrals do, again,
a fair amount of machinery must be developed before it can be used
efficiently. 

In this article, I shall (among other things)
try and give a simple treatment of both the
path integral and discrete phase integral methods. In order to orient this
discussion, however, it is first necessary to go back and ask what we mean
by spin tunneling to begin with. To this end, let us consider the toy
Hamiltonian
\beq
\ham_{\rm toy} = k_1 J_x^2 + k_2 J_y^2, \label{htoy}
\eeq
with $k_1 > k_2 > 0$. $\bJ$ is a dimensionless spin operator with the
usual commutation rules:
\beq
[J_i, J_j] = i\eps_{ijk} J_k.
\eeq
The first step is to understand the classical version of this problem.
What is the configuration space, the phase space, how are the classical
dynamics to be defined, and what are the classical states between which
tunneling will take place when quantum mechanics is turned on? To answer
these questions, we first note that a classical angular momentum (which
we also denote by $\bJ$) obeys the Poisson brackets
\beq
\{J_i, J_j\}_{\rm PB} = \eps_{ijk} J_k. \label{PB}
\eeq
Thus we can regard \eno{htoy} as a classical Hamiltonian (with the
dimensions of $\bJ$ and the $k$'s suitably readjusted), which defines
the dynamics of the vector $\bJ$ through the Poisson brackets 
(\ref{PB}):
\beq
{d \bJ \over dt} = \{\bJ, \ham\}_{\rm PB} 
      = - \bJ \times {\ptl\ham \over \ptl\bJ}. \lbl{eomJ}
\eeq
In fact, it is clear that (\ref{eomJ}) is a general prescription
applicable to any Hamiltonian that is a function of the components
of $\bJ$ only. An immediate consequence of \eno{eomJ} is that
\beq
{d \over dt} \bJ \cdot\bJ = - 2\bJ \cdot 
                \left( \bJ \times {\ptl\ham \over \ptl\bJ} \right) =0,
\eeq
so $|\bJ|$ is a constant of the motion, and configurations are completely
specified by giving the orientation $\Jhat = \bJ/|\bJ|$, and the
configuration space is the unit sphere. At the same time,
once $\Jhat$ is specified at any given time $t=0$, \eno{eomJ} allows us to
find it uniquely at any later time $t > 0$, as it is a first order
differential equation for $\Jhat$. Thus the unit sphere is a carrier
manifold in the language of modern classical mechanics, and it is also the
phase space. In other words, for spin, configuration and phase space are
one and the same.

If we interpret \eno{htoy} as a classical
Hamiltonian, then the classical energy
minima arise when $\bJ = \pm J\zhat$. These minima are degenerate. When the
problem is made quantum mechanical, these two classical states will appear as
two energy eigenstates with a small tunneling induced splitting. The problem
can be approached in two stages. First of all, classical states with definite
values of $\bJ$ go over into quantum mechanical states in which $\bJ$
automatically has a spread since the components of $\bJ$ do not commute.
The states with $\bJ = \pm J\zhat$ will acquire a zero point spread
$\avg{J_x^2} \sim \avg{J_y^2} \sim J$. (One can easily find these states
by solving and quantizing the equations of motion in the harmonic approximation.
This is completely equivalent to the Holstein-Primakoff procedure.) In the
second stage, tunneling mixes these states.

With this preamble, we can pose the basic question. What is the tunnel
splitting for a Hamiltonian such as (\ref{htoy}) which has two (or more)
degenerate
classical minima? In the course of answering this question, other questions
arise fairly quickly. For example, from a purely theoretical or mathematical
perspective, what are the associated wave functions? What are the splittings
between excited pairs of levels? The answers to these can be sought at varying
levels of rigour and quantitative accuracy. From the perspective of making
contact with experiments, one may want to know the influence of various
perturbations and ``dirt effects". Here, an important distinction needs to be
made between static and dynamic perturbations. For static perturbations, the
problem reduces to finding the matrix elements of various operators between
the tunneling states. The number of important operators in any system is
generally not large, so this type of question can be moved over into the
theory column in some sense. For time dependent perturbations,
two further subclasses need to be recognized. If the time dependence is of
the c-number type, i.e. due to a time-varying external field, the problem can
be reduced to either a standard NMR type, or to a Landau-Zener-St\"uckelberg
type. If the perturbation is due to other dynamical degrees of freedom, however,
the problem is much harder, and is in fact conceptually the same as that in
investigations of dissipation and decoherence in MQP. The range of possibilities
here is quite large in general, but in molecular magnetic systems one has the
advantage of knowing the relevant environmental degrees of freedom with a
high level of confidence, which greatly aids in theoretical modeling.

In this article, we shall largely be concerned with the theoretical aspects
of the problem. For specificity, we will focus on the \Fe8 system, but the
methodology is completely general. In Sec.~II, we will review the salient
features of the \Fe8 system, and discuss the results of the Wernsdorfer
and Sessoli experiment, with emphasis on the oscillatory field dependence
and vanishing of the tunnel splitting. We will compare these results with
numerical studies of the simplest model Hamiltonian for \Fe8, consider
the points where the splittings vanish in light of general quantum mechanical
theorems about degeneracy, and see that these points 
form a very rich pattern in the magnetic field space.

In Sec.~III, we will turn to the spin-coherent-state path integral approach.
These path integrals are much more delicate than the Feynman integral for
a massive particle, and the mathematical subtleties of the semi-classical
limit are still being researched. We will sidestep these points, and
concentrate on the Berry phase and the quenching condition for the special
case where the magnetic field is along the hard magnetic axis of the
molecule \cite{agepl93}.
This is the simplest case, and corresponds to a symmetric double well problem.
When the field is not along the hard or easy exes, the problem does not have
any symmetry, and instanton calculations, though possible, are quite involved.
More quantitative calculations can be done with comparatively greater ease
using a discrete phase integral (or WKB) method. We will discuss the basic
idea behind this method, and then show how this method can be used to find
tunnel splittings for \Fe8 for all orientations of magnetic
field \cite{agprl99,vf00,agprb00_1,agprb00_2}.

\section{The \Fe8 system}
\label{fe8sys}

\subsection{Summary of experimental facts and spin model}

The molecule \Fe8 (proper chemical formula:
[Fe$_8$O$_2$(OH)$_{12}$(tacn)$_6$]$^{8+}$) is magnetic, and%
\begin{figure}
\centerline{\psfig{figure=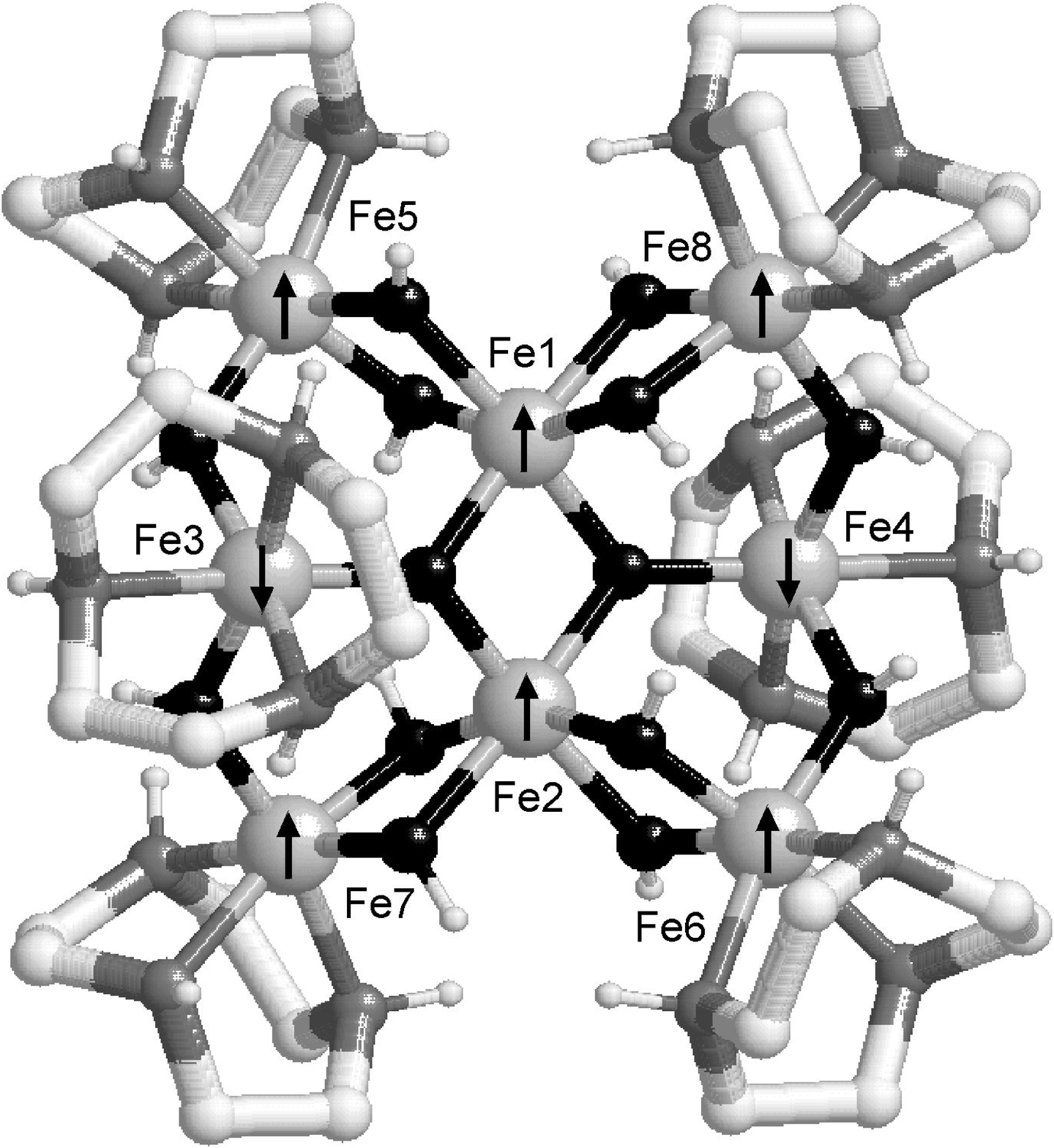,height=5cm}}
\caption{The \Fe8 molecule.}
\end{figure}
\noi forms good single
crystals. It has an approximate D$_2$ symmetry (see Fig. 2).
In its lowest
state, it is found to have a total spin of 10, arising from competing
antiferromagnetic interactions between the Fe$^{3+}$ ions within a molecule.
Spin-orbit and spin-spin interactions destroy complete rotational invariance,
and give rise to an anisotropy with respect to the crystal lattice
directions. A variety of experimental techniques (electron spin resonance,
ac susceptibility, magnetic relaxation, M\"ossbauer spectroscopy,
neutron scattering) indicates \cite{wpj84,d+93,bdg96,sop97,cam98}
that a single molecule can be described by the Hamiltonian
\beq
\ham_0 = k_1 J_x^2 + k_2 J_y^2 - g\mu_B \bJ\cdot\bH,
     \label{ham}
\eeq
with $J=10$, $k_1 \apx 0.33\ $K, and $k_2 \apx 0.22\ $K. In addition,
there are much weaker higher order anisotropies. The leading fourth order
anisotropy correction is
\beq
\ham_4 = -k_4 (J_+^4 + J_-^4), \lbl{ham4}
\eeq
with $k_4 \apx 2.9\times 10^{-5}\ $K \cite{ws99}.
The anisotropy energy is equivalent to a field of $\sim 2.5\ $T. The $g$
factor is very close to 2.

In writing the Hamiltonian for the eight coupled Fe$^{3+}$ spins in
terms of a single total spin $\bJ$ as we have done, the chief assumption
is that other spin multiplets are well separated in energy from the ground
multiplet. It is very hard to do first principles calculations of the
intramolecular,
interionic exchange parameters, and even when one can do this, it is very
hard to diagonalize the resulting Heisenberg exchange Hamiltonian. EPR
experiments do not show any evidence of other multiplets, and so while a
definite number is not known, it is not unreasonable to guess that other
multiplets will be separated from the $J = 10$ ground multiplet by at least
tens of Kelvin. Especially for the tunneling, the other multiplets can be
safely ignored.

Let us also briefly discuss environmental degrees of freedom which have
been left out of the description (\ref{ham}), and their
interaction with the spin. First, different molecules may interact with each
other. However, the \Fe8
molecule is very large, and in the solid, so is the intermolecular
separation. The primary interaction between molecules is dipolar (there
is no intermolecular exchange, in particular), and the dipolar field on any
molecule is about 100 Oe \cite{osp98},
much weaker than the intramolecular anisotropy field. Second, there is a
spin-phonon interaction, which as in the case of \Mn12, is responsible for
the moderate to high temperature magnetization relaxation \cite{bdg96},
and the dramatic hysteresis loop steps \cite{sop97} seen in this molecule too. 
At low temperatures, however, very few phonons are present in equilibrium, and
to a first approximation they may be neglected. This is especially so for
the tunneling phenomena we will consider in this article. Third, 
the atomic spins couple to the nuclear spins--the hyperfine interaction.
The biggest such coupling comes from magnetic nuclei in the magnetic species.
For iron, the only isotope with a magnetic nucleus is $^{57}$Fe, with a
natural abundance of 2-2.5\%. Thus, about 75\% of the molecules have no
nuclear spin in the magnetic species at all, and for those that do, the
hyperfine field seen by the electronic spin is about 10-100 Oe. If we average
the hyperfine field, and lump this along with the dipolar field in the
form of an inhomogeneous field, we get a distribution with a width of about
200 Oe \cite{osp98}, far smaller than the anisotropy field. (Alternatively,
we could say that the energy scales associated with dipolar plus hyperfine
interactions and the magnetic anisotropy are 0.2 K and 25 K, respectively.)
This approximation
omits the dynamical aspects of the nuclear spins, which are expected
to give rise to a small {\it unquenching\/} of the spin tunneling, i.e.,
make the transition probability nonzero. Thus, this aspect of the problem
is potentially important for a detailed understanding of the Wernsdorfer-%
Sessoli data, as is the distribution of dipolar and hyperfine fields.

In the rest of this article, we shall only study the pure spin problem
described by \eno{ham} and (\ref{ham4}). Further, for the most part,
we shall ignore the fourth order correction (\ref{ham4}).
This term is important in that even though
it is a small correction to the energy of any state, it significantly modifies
the location of the points in the magnetic field space where the splitting
vanishes. For the conceptual problem of understanding {\it why\/} we get
a vanishing splitting in the first place, however, it is sufficient to
study only $\ham_0$.  We shall discuss the effects of including
$\ham_4$ is Sec.~II$\,$D briefly.

\subsection{Tunneling states and the Landau-Zener-Stu\"ckelberg process}

If $\bH = 0$, $\ham_0$ is exactly the toy model of Sec.~I, and there are
two degenerate energy minima at $\pm\zhat$. If $\bH\ne0$, but still in the
{\it xy\/} plane, the classical minima move off the $\pm\zhat$ directions
but continue to be degenerate. One way to understand the tunneling between
these directions is to rewrite \eno{ham} after subtracting out a constant
term $k_2\bJ\cdot\bJ = k_2 J(J+1)$, and explicitly putting $H_z = 0$. This
yields
\beq
\ham = -k_2 J_z^2 + (k_1 - k_2)J_x^2
        -g \mu_B (J_x H_x + J_y H_y).
\eeq
Let us now regard the last three terms as perturbations that give rise
to transitions between various Zeeman levels or eigenstates of $J_z$.
As usual, we denote the $J_z$ value by $m$. The $J_x^2$ term gives rise to
$\Dta m = 2$ transitions, and thus mixes $m=-10$ with $m = +10$ via the
$-8$, $-6$, $\ldots$, $+8$ states. The $H_x$ and $H_y$ terms give rise
to $\Dta m= 1$ transitions, and mix $m=-10$ with $+10$ via all intermediate
levels $-9$ to $+9$ (see Fig.~3a). This picture allows us to think of
spin tunneling in direct analogy with a particle tunneling through an
energy barrier. It is further obvious that if we also apply a field along
the $z$ direction so as to tune the energies of the $-10$ and $+9$ states
to resonance, we can also think of tunneling between these states. More
generally, we can consider tunneling between a state with $m=m_i$ on the
negative $m$ side and $m = m_f$ on the positive $m$ side.
\begin{figure}
\centerline{\psfig{figure=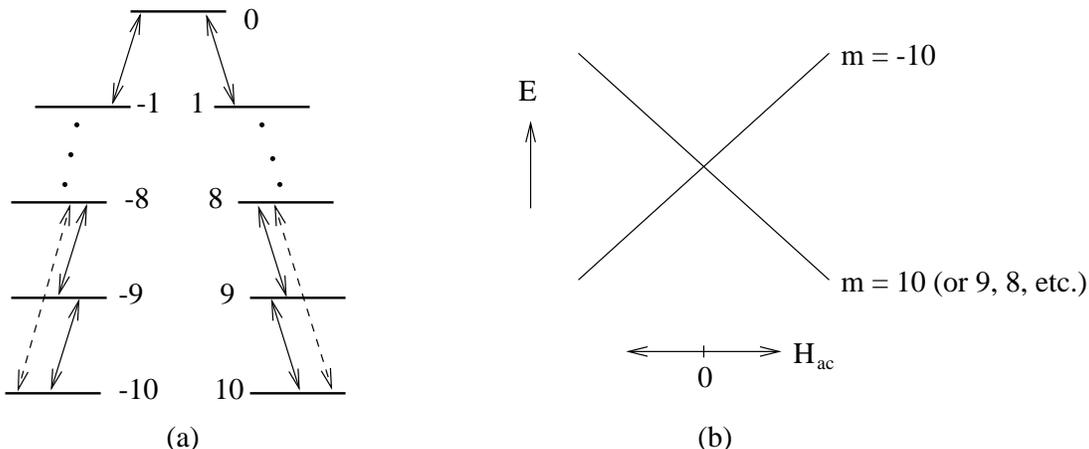}}
\caption{(a) Zeeman levels of \Fe8, showing $\Dta m =1$ (solid lines)
and $\Dta m = 2$ (dashed lines) transitions. (b) The Landau-Zener-%
St\"uckelberg process.}
\end{figure}

At this point, it is appropriate to ask if the tunnel splitting is
big enough to measure experimentally. Although one of the main goals
of this article is to understand
such splittings analytically, let us temporarily
cheat and diagonalize the $21 \times 21$ Hamiltonian matrix numerically
with the experimental numbers for $k_1$ and $k_2$. When this is done, we
find that the splitting, $\Dta \sim 10^{-8}$ K. We now recall that
the bias or departure from degeneracy between the $m=\pm 10$ states
caused by the dipolar and hyperfine fields, which we denote by $\eps$,
is about 0.1 K,
which is enormous compared to $\Dta$. Hence, left to itself, a spin
will have almost no chance of tunneling at all. To see this, consider
a two level Hamiltonian
\beq
\left(
\begin{array}{cc}
\eps & \Dta \\ -\Dta & -\eps
\end{array}
\right).
\eeq
It is now easy to verify that if we start at $t=0$ in the $+\eps$
level, the probability of finding the system in the $-\eps$ level at a later
time oscillates with a frequency, $2(\Dta^2 + \eps^2)^{1/2}$,
and an amplitude $\Dta^2/(\Dta^2 + \eps^2)$. If $\eps \gg \Dta$, the
transition probability is always very small. The same general conclusions
apply to any two levels $m_i$ and $m_f$, not just the $\pm 10$ states.

Wernsdorfer and Sessoli solve this difficulty by making use of the
Landau-Zener-St\"uckelberg (LZS) mechanism to induce tunneling transitions%
\cite{lzs32}.
A dc longitudinal field $H_z$ is applied so as to bring $m_i$ and $m_f$
into approximate resonance. They then
apply an additional ac {\it longitudinal} magnetic field $H_z(t)$
in the form of a triangular wave. Denoting the amplitude of this wave
by $H_0$ and the time period by $\tau$, we have
$\dot H = |dH_z/dt| = 4H_0/\tau$. Now, as $H_z$ changes with time, the
energies of the $-10$ and $+10$ levels will move in opposite directions, and
at some point in the cycle they will cross. (See Fig. 3b.)
This gives rise to what is known
as the LZS process. Basically, in the vicinity of the crossing,
the energy bias goes to zero, and there is an appreciable chance for the spin
to tunnel. The probability for a transition during one crossing is given
by \cite{llqm}
\beq
P = 1 - e^{-\gam},
\eeq
where
\beq
\gam = {2\pi \Dta^2 \over (g\mu_B\hbar) \Dta m \dot H},
\eeq and $\Dta m = |m_f - m_i|$ is the change in $m$ in the transition.
The quantity
$\gam$ is known as the adiabaticity parameter, and in the limit where the 
crossing is passed rapidly, i.e., $\dot H$ is large, $\gam \ll 1$, and
$P \apx \gam$. We can further assume that because the stray fields are
fluctuating, the phase of the molecule will be randomized and uncorrelated
between successive crossings. If $H_0$ is large enough to overcome the range
of dipolar biases (but not so large as to make more than one Zeeman level on
the positive side cross the level on the negative side), every molecule in
the sample will undergo a crossing at some point in the cycle. Since there
are $2/\tau$ crossings per unit time, we obtain a transition probability
per unit time for $m_i \tofro m_f$ transitions, given by
\beq
\Gam_{\rm LZS} = {2 \over \tau}\gam 
           \apx {\pi \Dta^2 \over (g\mu_B\hbar) \Dta m H_0}
                           \quad (\gam \ll 1).
           \lbl{ratelzs}
\eeq

Wernsdorfer and Sessoli first saturate the sample in a large longitudinal
field. This field is then removed and the ac field is applied, inducing LZS
transitions, and causing a relaxation of the magnetization of the sample.
By measuring the rate of this relaxation, one can obtain $\Gam_{\rm LZS}$, from
which one can in turn infer $\Dta$ using \eno{ratelzs}
and the experimental value
of $H_0$. An important check on the consistency of the LZS interpretation
implied by \eno{ratelzs} is that the relaxation rate should be independent
of the sweeping rate $\dot H$. Experimentally, this is found to be true for
$\dot H$ ranging from 1 mT/s to 1 T/s.

\subsection{Oscillatory tunnel splittings, the von Neumann-Wigner theorem,
            and diabolical points}

It is apparent that the LZS measurements can be carried out in the presence
of a transverse dc field (in the {\it xy\/} plane) $H_\perp$, so that
$\Dta$ can be measured as a function of $H_\perp$. Naively, we expect
that since increasing $H_\perp$ decreases both the energy barrier and the
angle through which the spin must tunnel, $\Dta$ will increase monotonically
with $H_\perp$. What is actually seen%
\begin{figure}
\centerline{\psfig{figure=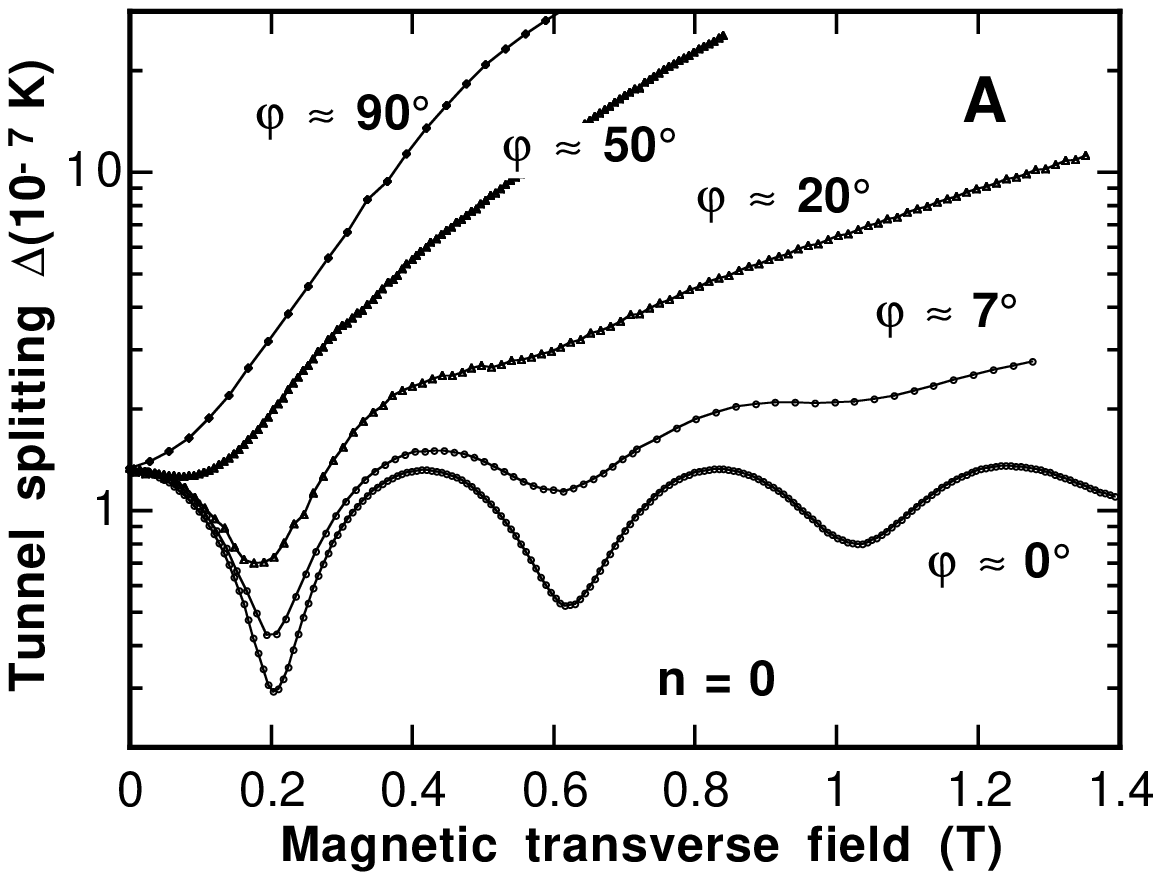,height=5cm,width=0.46\linewidth}
\psfig{figure=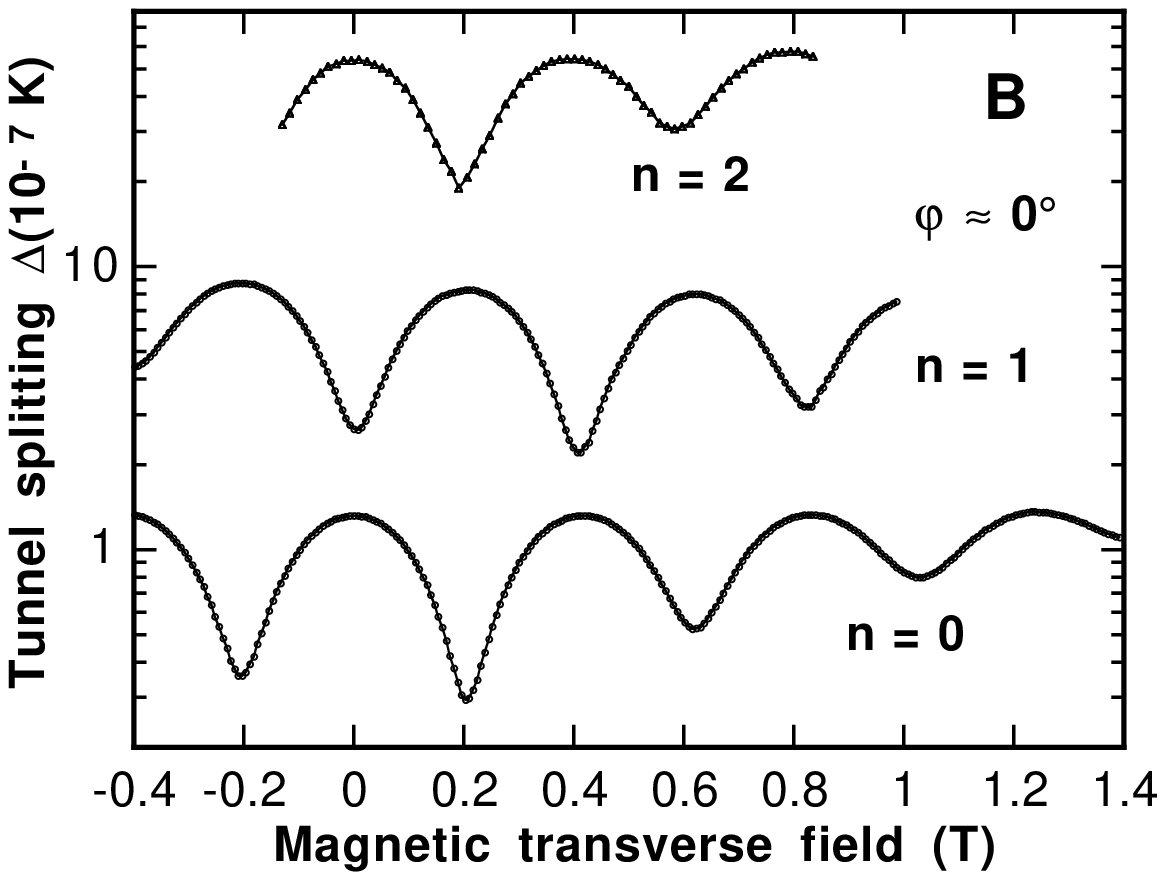,height=5cm,width=0.46\linewidth}}
\caption{Measured splittings [12] for \Fe8 for (a) $-10 \tofro 10$
transitions for various orientations of $\bH$ in the {\it xy\/} plane,
and (b) for $\bH\|\xhat$ between the states $m =-10$ and $m= 10 -n$.}
\end{figure}
\noi experimentally is rather different (see Fig.~4).
When $\bH \| \yhat$ ($\phi = 90^{\circ}$), the behavior is indeed monotonic,
but when $\bH \| \xhat$, one finds that $\Dta$ oscillates with $H_x$.

We have already stated in Sec.~I that this oscillation can be understood in
terms of a Berry phase in the spin path integral. Since the vanishing of
$\Dta$ means that two energy levels of the system are exactly degenerate, it
is useful to examine this result in general quantum mechanical terms, and
from the perspective of rigorous results about when such degeneracies can and
cannot occur. Before we turn to this, however, it is useful to look at the
results of a numerical diagonalization of the Hamiltonian (\ref{ham}). These
numerical data reveal a number of other properties, some of which are
special to the form (\ref{ham}), but others are general.  In Fig.~5, we show 
the results of numerical calculation of the energies as a function of
$H_x$, for%
\begin{figure}
\centerline{\psfig{figure=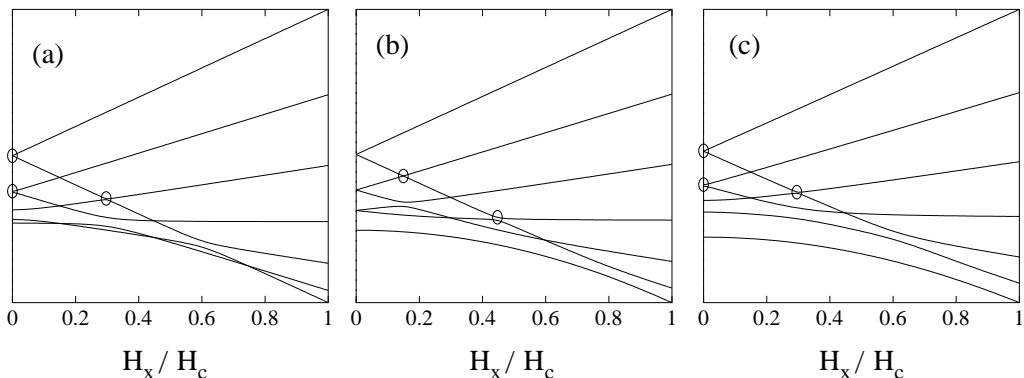,height=5cm}}
\caption{Spectrum of the Hamiltonian (2.1) for $J=3$, as a function
of $H_x/H_c$, with $H_c =2k_1 J/g\mu_B$. $H_z/H_c = 0$,
$0.07454$, and $0.1491$ in (a), (b), and (c), respectively. The small
ovals indicate narrowly avoided anticrossings that appear to be
crossings on the resolution of this figure.}
\end{figure}
\noi 
$J=3$, for three different values of $H_z$. In all three cases,
$H_y = 0$. In part (a), $H_z = 0$, and problem is like a symmetric double
well. We see that the lowest two energy level curves cross
a number of times. These crossings correspond to the curve marked $n=0$,
i.e. to $-10 \tofro +10$ transitions in Fig. 4b. In addition,
Fig.~5a also shows a number of crossings of higher energy levels.
These crossings
are difficult to see directly but their presence has been inferred indirectly
by studying the temperature dependence of the shape of the minima in $H_x$
dependence of the transition rate \cite{wsc00}.

In Fig.~5b, $H_z$ has a specific non-zero value, chosen so that (ignoring
tunneling), the first excited state in the deeper well is degenerate
with the lowest state in the shallower well.  The problem is no longer
symmetric, and one of the classical minima is lower than the other.
Correspondingly, we see that the lowest quantum mechanical state is always
non degenerate. However, we see from the figure that the second and third
energy levels cross a number of times.  These crossings correspond to the
curve marked $n=1$, i.e. to $-10 \tofro +9$ transitions in Fig.~4b. 
As seen in the experiments, the crossings in Fig.~5b are shifted by
half a period from those in Fig.~5a. Further, as in part (a), we see
crossings between yet higher energy levels (the fourth and fifth, e.g.).

This pattern continues as $H_z$ is increased still further (Fig.~5c). Now the
lowest two levels in the deeper well are nondegenerate, and the lowest crossings
are between levels 3 and 4. Compared to Fig.~5b, these crossings
are shifted by yet another half-period, just as seen experimentally.
Again, there are crossings between higher pairs
of levels.

Let us now recall the conditions under which energy levels of a quantum
mechanical system may intersect under variation of a parameter. This is governed
by the von Neumann-Wigner theorem \cite{vnw29}, which states 
that as a single parameter in a Hamiltonian is varied, an intersection
of two levels is infinitely unlikely, and that level repulsion is the rule.
It is useful to review the argument behind this theorem. Let the
energies of levels in question be $E_1$ and $E_2$, which we suppose to be far
from all other levels. Under an incremental perturbation $V$, the secular
matrix is
\beq
\left(
    \begin{array}{cc}
        E_1 + V_{11} & V_{12} \\
        V_{21}       & E_2 + V_{22}
    \end{array}
\right),    \label{sec}
\eeq
with $V_{21} = V^*_{12}$.
The difference between the eigenvalues of this matrix is given by
\beq
[(E_1- E_2 + V_{11} - V_{22})^2 + 4 |V_{12}|^2]^{1/2}, \label{edif}
\eeq
which vanishes only if
\beq
E_1 + V_{11} = E_2 + V_{22}, \quad\quad V_{12} = V^*_{12} = 0. \label{cond}
\eeq
Hence, for a general Hermitean matrix, three conditions must be satisfied for
a degeneracy, which in general requires at least three tunable parameters.
If the matrix is real and symmetic, the number of conditions and tunable
parameters is reduced to two \cite{via78}. 
Degeneracies of the latter type, i.e. those obtained by tuning more than
one parameter are known as {\it diabolical\/}, or in older
terminology, as {\it conical intersections\/} \cite{bw84,hlh63}. The reason
for this terminology is that if we denote the experimentally controllable
parameters by $x$ and $y$, and define these to be zero at the intersection, then, in its vicinity, we may expand the various contributions to
Eq.~(\ref{cond}) as
\bea
E_1 + V_{11} - E_2 + V_{22} &\apx& a_x x + a_y y, \nnu\\
V_{12} &\apx& b_x x + b_y y, \nnu
\eea
and the energy surface is given by
\beq
E = {\rm constant} \pm \bigl[ (a_x x+ a_y y)^2 
                             + (b_x x + b_y y)^2 \bigr]^{1/2}.
\eeq
This is an elliptic double cone in the {\it xy\/} plane, resembling in
shape an Italian toy called the {\it diavolo\/}.

An exception to the no-crossing rule occurs when the
Hamiltonian has some symmetry, when
levels transforming differently under this symmetry {\it can} intersect.
In the \Fe8 moel, the Hamiltonian is invariant under 180$^\circ$ rotations
about $\bH$ when $\bH\|\xhat$ or $\bH\|\zhat$, so there is such a symmetry,
and states which are even or odd under the relevant operation can
cross \cite{agprb95}. When $H_x$ and $H_z$ are both non-zero, however,
there is no geometrical symmetry, and the crossings in Fig.~5(b) and (c),
and the corresponding minima seen by Wernsdorfer and Sessoli, are non trivial
instances of diabolical points. Note that the conditions of the theorem
cited above are met no matter how one counts the parameters. If we regard
the system as having two parameters, $H_x$ and $H_z$, then the Hamiltonian
can be chosen to be real by using the standard representation of the angular
momentum matrices. All three angular momentum matrices $J_x$, $J_y$, and
$J_z$, cannot be made real simultaneously, however, so if we regard the
parameter space as three dimensional $(H_x, H_y, H_z)$, the Hamiltonian
is complex.  In either case, a degeneracy can only occur at an isolated
point. Thus, viewed either in the larger $H_x$--$H_z$ plane, or in the full
three-dimensional space of magnetic fields $\bH$, {\it all} points of
degeneracy, including those on the $H_x$ and $H_z$ axes, are diabolical.

It should be noted that in real physical systems, even when more than parameter
can be varied, diabolical points are quite rare. \Fe8 is remarkable in having
such a rich pattern of intersections. Although, as already mentioned, the
locations of the diabolical points depend sensitively on the presence of the
fourth order term $\ham_4$, it is of considerable theoretical interest that
for the simpler model where the Hamiltonian is taken to be just $\ham_0$,
a number of results can be proved exactly \cite{kg00}. (Surprisingly, the
semiclassical analyses \cite{agepl93,agprl99,vf00,agprb00_1,agprb00_2}
seem to capture these results exactly at leading order in $1/J$.)
The first of these is for
the location of the diabolical points. The point where the $\ell'$th
level in
the negative $J_z$ well (with $\ell'=0$ being the lowest level) and the
$\ell''$th level in the positive one are degenerate, is at $H_y = 0$, and
\bea
{H_z(\ell',\ell'') \over H_c}
    &=& {\sla (\ell'' - \ell') \over 2 J}
                 \label{plhz} \\
{H_x(\ell',\ell'') \over H_c}
   &=& {\sqrt{1-\lam} \over J}
        \left[ J - n - \tshf (\ell' + \ell'' + 1) \right],
             \label{plhx}
\eea
with $n = 0, 1, \ldots, 2J - (\ell' + \ell'' + 1)$. Here,
$\lam = k_2/k_1$, and $H_c = 2 k_1 J/g\mu_B$. Thus, the diabolical points lie
on a perfect centered rectangular lattice in the $H_x$-$H_z$ plane
(Fig.~6).
\begin{figure}
\centerline{\psfig{figure=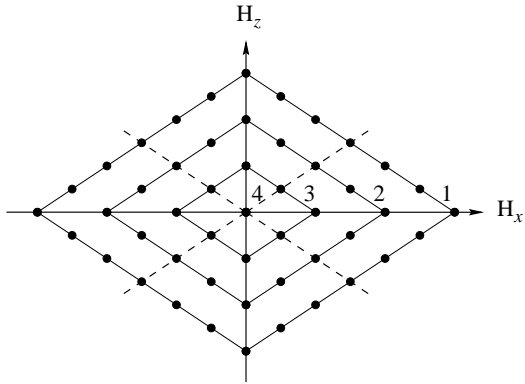,height=5cm}}
\caption{Diabolical points for the Hamiltonian \ref{ham} for $J=7/2$.
Many of the points are multiply diabolical, i.e. correspond to more than
one pair of simultaneously degenerate levels. This multiplicity is the
same for all points on a rhombus and is as shown.}
\end{figure}

Secondly, many of the points are {\it multiply}
diabolical, i.e., more than one pair of levels is simultaneously
degenerate. It is easily shown that the multiplicity is as indicated
in Fig.~6: If we arrange the points into concentric rhombi, those on the
outermost rhombus are singly diabolical (i.e., there
is only one pair of degenerate states), those on the next rhombus
are doubly diabolical (two pairs of degenerate states), and so on.

These facts hint very strongly at the presence of a higher dynamical symmetry,
i.e., an additional conserved quantity. This symmetry has not been found so far,
but knowing it would be a tremendous advance even for real \Fe8, as it would
be only weakly broken. For the same reason, it would be very useful to know
the exact wavefunctions at the diabolical points, since that would enable one
to study corrections and perturbations more systematically.

\subsection{Influence of higher order anisotropy perturbations}
\label{higher}

It is of some interest to ask what happens to the diabolical points when
the fourth-order perturbation (\ref{ham4}) is included. Let us first
investigate this question without making use of the specific form of the
perturbation, from the general point of view of enlarging the parameter
space of the Hamiltonian. Keeping $H_y = 0$, we think of our Hamiltonian as
depending on three parameters, $H_x$, $H_z$, and $k_4$. The general argument
about the codimension of a degeneracy \cite{via78} implies that in the
three-dimensional $(H_x,H_z,k_4)$ space, a diabolical point%
\begin{figure}
\centerline{\psfig{figure=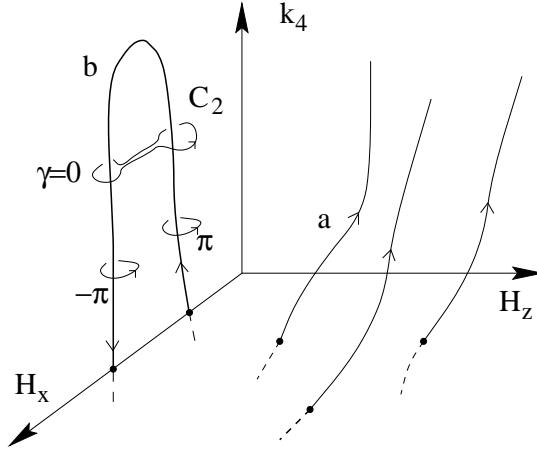,height=6cm}}
\caption{Trajectories of diabolical points under the influence
of a perturbation $k_4$. The values $\pm\pi$ and $0$ are the Berry
phases associated with the adjacent contours.}
\end{figure}
\noi in the
initial $(H_x,H_z)$ plane turns into a line (see Fig.~7).
In the figure, we show the two kinds of possible behavior that are
permitted by this theorem, and that are topologically allowed. The first, as
in the line marked `a', shows a diabolical point that continues on
indefinitely. The second possible behavior is shown in the line marked
`b'. What appear to
be distinct diabolical points in the $k_4 = 0$ plane,
lie, in fact, on the same diabolical line in the three-dimensional space.
More generally, diabolical lines can formed closed loops, but cannot
terminate abruptly.

A second way of viewing this matter is provided by Berry's phase
\cite{mvb84}. Suppose that at some value of $k_4$, two states
$\ket{\psi_a(k_4,\bH)}$ and $\ket{\psi_b(k_4,\bH)}$ are degenerate at
$\bH = (H_{x0},0,H_{z0}) \equiv \bH_0$. Let $C$ be a small closed contour
in the $H_x$-$H_z$ plane encircling the point $\bH_0$.
Berry's phase is given by
\beq
\gam(C) = i \oint_C
   \tran{\psi_a(k_4,\bH)}{\nabla_{\bH}\psi_a(k_4,\bH)}\cdot d\bH,
         \label{berph}
\eeq
where $\bH$ is now the two-dimensional vector $(H_x,H_z)$.
As shown by Berry, $\gam(C) = \pm\pi$
if $C$ encloses a true diabolical point, and $\gam(C) = 0$ if the
two states merely approach each other very closely without ever being
degenerate. [Actually, since our Hamiltonian is real and the parameter
space $(H_x,H_z)$ is two-dimensional, we really only need the weaker
and older result due to Herzberg and Longuet-Higgins \cite{hlh63}, which
states that the states $\ket{\psi_a}$ and $\ket{\psi_b}$ change sign upon
encircling the degeneracy: $e^{i\gam(C)} = -1$. This sign change test is
an efficient way of searching for diabolical points numerically.]
Now suppose that any of the three parameters is changed slightly. Since
the perturbation corresponding to this change is non-singular,
$\ket{\psi_a(k_4,\bH)}$ is a smooth function of $k_4$ and $\bH$. It
follows that if $k_4$ varies continuously, the integrand of
\eno{berph} can not change discontinuously. Hence, for small
enough $\dta k_4$, the phase $\gam(C)$
must continue to be what it was for $k_4 = 0$, $+\pi$, say,
implying that $C$ continues to encircle a degeneracy if it did so
at $k_4 = 0$.

From this point of view, the behavior `b' in Fig.~7 can only arise if
$\gam(C)$ has opposite values for the contours encircling the two
diabolical points at low values of $k_4$. (Naturally, both contours
must have the same sense.) The Berry phase for a contour $C_2$
encircling both points is then $0$, and it is then possible that for
$k_4$ exceeding some value $k_4^*$, we can shrink $C_2$ to zero,
without encountering any singularity. It is obvious that this can happen
only if the two diabolical points annihilate each other at $k_4^*$.
Pictorially, we can imagine ``slipping" the contour $C_2$ off the
diabolical line by moving it above the hairpin bend in the figure.

In Fig.~8, we show the results of a numerical calculation (performed
by E. Ke\c{c}ecio\u{g}lu), for $J=10$, using the experimental values
of $k_1$ and $k_2$ pertinent to \Fe8. The figure shows%
\begin{figure}
\centerline{\psfig{figure=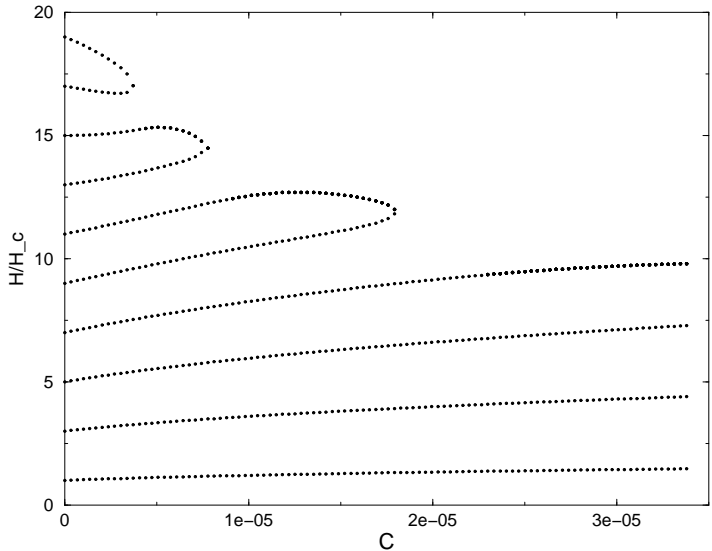,height=8cm,angle=270}}
\caption{Degeneracy fields for lowest two energy levels with $\bH\|\xhat$
as a function of fourth order anisotropy. The quantity $C$ is
identical to $k_4$, and is given in Kelvin.} 
\end{figure}
\noi how the diabolical
points on the line $H_z = 0$ that correspond to the degeneracy of the
lowest two energy levels, move under the influence of the fourth order
term ($C \equiv k_4$). We see three examples of a hairpin bend where
two diabolical points annihilate. There are two points worthy of
comment. First, the points that were on the surface $H_z = 0$ when
$k_4 = 0$, continue to be on the same surface when $k_4 \ne 0$. This can
be seen as a consequence of symmetry. When $H_y = H_z =0$,
$\ham_0 + \ham_4$ continues to be invariant under a $180^{\circ}$
rotation about $\xhat$. Thus, under a change in $k_4$, levels which
cross at $k_4 = 0$ because they have different signs under this symmetry
operation, can continue to cross only if we continue to have
$H_y = H_z = 0$. Second, by the time we get up to
$k_4 = 3\times 10^{-5}$~K, only four diabolical points survive on the
$H_x$ axis, and the spacing between them is nearly 50\% greater than
the period at $k_4 = 0$. Both these facts are in agreement with the
experimental data. In fact, it was by using
this argument in reverse,
i.e., by fitting to the observed spacing that Wernsdorfer and Sessoli
deduced the value of $k_4$ quoted in Sec.~II$\,$A. Since the pattern
and location of diabolical points is sensitively influenced by $k_4$,
this would appear to be a more reliable method of finding higher order
anisotropy coefficients than direct EPR spectroscopy.

\section{Instanton calculation of tunnel splittings}
\label{instanton}

In this section, we will turn to the instanton method \cite{jl67,sc77+}
of calculating the ground state tunnel
splitting, which is based on spin coherent state path integrals. Our main
goal is to understand the quenching effect fairly rapidly, so we will skip
lightly over the subtleties in the spin path integral and the semiclassical
limit, which are far more vexing than those for massive particles. (We will,
however, briefly describe what these subtleties pertain to in subsection E.)

\subsection{What to calculate: the imaginary time propagator}

We use the overcomplete basis of spin coherent states $\{\ket\nhat\}$ to
describe our spin. The state $\ket\nhat$ has maximal spin projection along the
unit vector $\nhat$:
\beq
 \bJ\cdot\nhat \ket{\nhat} = J \ket{\nhat}.
\eeq
We now consider the imaginary time transition amplitude
\beq
U_{21} = \mel{\nhat_2}{e^{-\ham T}}{\nhat_1},
      \lbl{U21}
\eeq
where $\nhat_{1,2}$ are the minima of the classical energy
\beq
E_{\rm cl}(\nhat) = \mel{\nhat}{\ham}{\nhat}.
\eeq

Let us now denote the two lowest energy eigenstates by $\ket{\psi_{\pm}}$,
and their energies by $E_\pm$. where
\beq
E_{\pm} = E_{\rm av} \pm {\tshf}\Dta,
\eeq
with $E_{\rm av}$ being the average, and $\Dta$ the splitting as defined
earlier. We expect that since these states are tunnel split states, they
can be well approximated as linear combinations of states
$\ket{\psi_{1,2}}$ that are well localized around the corresponding
directions $\nhat_{1,2}$. In other words,
\beq
\kpm = {1 \by \sqrt2} \left( \kp1 \pm \kp2 \right),
         \lbl{psipm}
\eeq
with
\bea
\tran{\nhat_i}{\psi_i} &\equiv& a_i \ne 0, \\
\tran{\nhat_i}{\psi_j} &\simeq& 0, \quad (i \ne j).
\eea
Note that we can always choose the phases of $\kp i$ so that \eno{psipm}
is correct as written.

Next, we expand the amplitude $U_{21}$ in terms of the complete set of
energy eigenstates. As $T \to\infty$, only the lowest two states will
contribute, and we get
\bea
U_{21} &\apx& \tran{\nhat_2}{\psi_+} \tran{\psi_+}{\nhat_1}e^{-E_+ T} 
           + \tran{\nhat_2}{\psi_-} \tran{\psi_-}{\nhat_1}e^{-E_- T} \nnu\\
       &=& \tshf a_2 a_1^* e^{-E_+ T} - \tshf a_2 a_1^* e^{-E_- T}  \nnu\\
       &=&  a_2 a_1^* e^{-E_{\rm av} T}  \sinh(\Dta T).
          \lbl{U21exp}
\eea

\subsection{The spin-coherent-state path integral}

Our goal is to calculate $U_{21}$ via path integrals,
and compare with \eno{U21exp} in order to obtain $\Dta$. The essential factor
that one seeks to capture is $\sinh(\Dta T)$. As in the case of massive
particles \cite{rsqm}, we write the amplitude as a sum over paths on the
unit sphere, weighting each path by the exponential of the action for
that path:
\beq
U_{21} = \int_{\nhat_1}^{\nhat_2} [d\nhat]\, e^{-S[\ntau]}.
\eeq
The paths $\ntau$ all run from $\nhat_1$ at $-T/2$ to $\nhat_2$ at
$T/2$, and $S[\ntau]$ is the imaginary time or Euclidean action for the path.
(This is why the exponent is $-S$ rather than $iS/\hbar$.)
The novel aspects for spin lie in the nature of this action, which
is given by
\beq
S[\ntau] = iJ \ant + \int_{-T/2}^{T/2} \Ecl[\ntau] d\tau.
  \lbl{Sofn}
\eeq
Instead of deriving this result, we shall show that it is correct
by checking that its variation leads to the classical equation of motion.
The term $\ant$ is the kinetic term, and has the mathematical structure of
a Berry phase. (The same term is responsible for the Haldane gap in
one-dimensional antiferromagnets \cite{dh83+,ef91}.)
One of the key properties of such a term
is that it can not be made manifestly gauge invariant, and this fact has led
to misstatements in the literature in which a representation arising from
a particular gauge choice, and the coordinate singularities of spherical polar
coordinates are said to be responsible for its topological properties. In
order to avoid these pitfalls, we write it as
\beq
\ant = \oint_{\crv = \ntau - \nhat_R} d\Om, \lbl{Asurf}
\eeq
by which we mean that $\ant$ is the solid angle enclosed by the closed curve
formed by the path $\ntau$ and a reference path $\nhat_R$ (taken backwards)
running from $\nhat_1$ to $\nhat_2$ (Fig.~9). The reference path $\nhat_R$
is arbitrary, but must be the same for all $\ntau$ in the path integral.
Its choice is equivalent to fixing the gauge. However, since we have
defined $\are$ as a geometrical quantity, an area, it does not depend on how
we choose coordinates on the unit sphere, and%
\begin{figure}
\centerline{\psfig{file=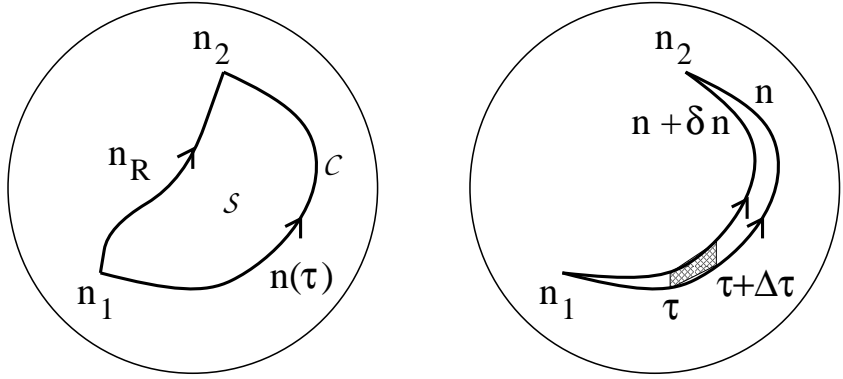,height=5cm}}
\caption{The kinetic term in the spin action, and its variation}
\end{figure}
\noi it is obviously nonsingular.

The actual evaluation of $\are$ for any given closed path is often more simply
done by using Stokes's theorem to transform it to a line integral. Using
notation borrowed from electromagnetism, let us write
$d\Om = \bB\cdot\nhat ds$, where $\bB(\nhat) = \nhat$, and $ds$ is an area
element. The line integral is $\int\bA \cdot d\nhat$, with $\bB = \nabla\times\bA$.
Since $\bB = \nhat$, it is the magnetic field of a monopole. It is known 
that if we try and represent a monopole field in terms of a vector potential
$\bA$, then $\bA$ must have a singularity somewhere. If this singularity is
concentrated into a Dirac string at the south pole, we can write
\beq
\ant = \oint_{\crv} \bigl[ 1 - \cos\tta(\phi) \bigr] \,d\phi,
   \lbl{Aline}
\eeq
where the curve $\crv$ is regarded as being parametrized by $\phi$. This
formula is correct as long as $\crv$ does not pass through the south pole,
and provided one increments or decrements $\phi$ by $2\pi$ every time one crosses
the date line.

Let us now obtain the classical equations of motion by varying the action.
Consider the kinetic term first. Suppose we vary the path $\ntau$.
The variation $\dta\are$ is given by the area of the thin sliver on the
sphere enclosed between the curves $\ntau$ to $\ntau +\dta\ntau$ (Fig.~9).  
The part of this area due to the segments between $\tau$ and $\tau + \dta\tau$
is given by
\bea
\Dta(\dta\are) &=& \Bigl[ \dta\ntau \times
                           [ \nhat(\tau + \Dta\tau) - \ntau] \Bigr]
                     \cdot\ntau \nnu\\
          &=& \left( \dta\ntau \times {d\nhat \by d\tau} \right)
                  \cdot \ntau \Dta\tau.
\eea
Adding up the contributions from all the segments, we find the total change
\beq
\dta\are = \int \dta\ntau \cdot
            \left({d\nhat \by d\tau} \times \ntau \right) \, d\tau.
\eeq
The variation of the second term in \eno{Sofn} is
\beq
\int \dta\ntau \cdot {\ptl \Ecl \by \ptl\nhat} d\tau.
\eeq
Thus the variation of $S$ can be written as an integral of the form
$\int\dta\ntau\cdot X$ where $X$ is something depending on $\nhat$ and $\Ecl$.
The condition for $\dta S$ to vanish is thus $X=0$, or
\beq
i J {d\nhat \by d\tau} \times \ntau + {\ptl \Ecl \by \ptl\nhat} = 0.
\eeq
Taking the cross product of this equation with $\nhat$, and making use of the
fact that $\nhat\cdot(d\nhat/d\tau) =0$, we get
\beq
iJ \dndt = - \left( \nhat \times \dedn \right). \lbl{eomJ2}
\eeq
This equation is exactly what we would get from \eno{eomJ} with $\bJ = J\nhat$
and the Wick rotation $t \to -i\tau$. In other words, it is 
the imaginary time equation for Larmor precession in the effective
magnetic field $\ptl\Ecl/\ptl\nhat$. (It is also called the Landau-Lishitz
equation in magnetism.)

One consequence of \eno{eomJ2} is that $\Ecl$ is conserved along the
classical path. For,
\bea
{d\Ecl \by d\tau} &=& \dedn\cdot\dndt \nnu\\
                  &=& {i \by J} \dedn \cdot \left(
                        \nhat \times \dedn \right) \nnu\\
                  &=& 0.
\eea
[Another way to see this is that if \eno{eomJ2} is written in
terms of spherical polar coordinates, it will be seen to have a
Hamiltonian structure with $\phi$ and $J\cos\tta$ as canonically
conjugate variables, and $\Ecl$ as the Hamiltonian.]

\subsection{How to calculate the propagator: instantons}

The path integral (\ref{U21}) is not easy to evaluate. In the
$T \to\infty$ limit, however, we can use the approximation of steepest
descents. The dominant paths, known as instantons,
are just those that minimize the action, i.e. they are solutions
to the classical equations of motion. The simplest such paths consist of
a single transit fro $\nhat_1$ to $\nhat_2$. If the scale over which $\Ecl$
varies is $V$, then it follows from \eno{eomJ2} that the time scale for
this transit is $\tau_0 \sim J/V$. For the toy Hamiltonian
(\ref{htoy}), e.g., this
time scale is $J^{-1} (k_1 k_2)^{-1/2}$. Since $T \to \infty$, it follows that
the spin spends most of its time near the end points $\nhat_{1,2}$, and the
actual transit takes place in a very short time interval. (Hence the name
instanton.) Further, since the equation of motion is autonomous, i.e., does not
depend on $\tau$ explicitly, in the $T\to \infty$ limit, it is clear that
a translation of the center of the instanton yields an equally good classical
path. Once this is realized, it is not difficult to see that one can
have multi-instanton solutions, in which the path goes between $\nhat_1$
and $\nhat_2$ several times, with the centers of the instantons being
widely separated on the time scale $\tau_0$.  When all the contributions
to $U_{21}$ are evaluated in this way, one finds that the $n$-instanton
terms give a contribution proportional to $T^n$, and the full series is
that of a $\sinh$ \cite{sc77+}. The $\Dta$ which is obtained in this way
can be written as
\beq
\Dta = D \exp(-S_{\rm inst}),
\eeq
where $S_{\rm inst}$ is the action for a {\it single instanton} path, and
$D$ is a prefactor arising from doing the path-integral over small fluctuations
about the instanton trajectory. If more than one instanton exists, we must
add together the corresponding contributions from all of them.

It is important to note that since $\Ecl$ is conserved along the instantons,
and since $\nhat_1$ and $\nhat_2$ are minima of $\Ecl$, there can not be any
real path $\ntau$ conecting them. The only solution is to allow $\nhat$ to become
complex. Correspondingly, the area $\are$ must be defined on the complexified
unit sphere. We can take $\Ecl$ to be zero for an instanton by adjusting the
zero of energy, so the action is just $i J \are$. The problem is thus reduced
to finding the instantons and the corresponding area. This is extremely
simple, however. Since we can write $\ant$ as the line integral (\ref{Aline}),
we do not
need to find the actual time dependence of the instanton path, and it suffices
to find the orbit on the unit sphere, i.e., $\tta$ in terms of $\phi$, which
can be done from energy conservation alone.

\subsection{Application to \Fe8}

Let us illustrate this procedure for \Fe8, with the Hamiltonian $\ham_0$ for
the case where $\bH\|\xhat$.
If we choose the polar axis to be $\xhat$ (not $\zhat$), and measure the azimuthal
angle in the {\it yz\/} plane from $\yhat$, then we can write
\beq
\Ecl(\tta,\phi) = k_1 J^2 (\cos\tta - \cos\tta_0)^2
                  + k_2 J^2 \sin^2\tta\sin^2\phi.
\eeq
We have defined $\cos\tta_0 = H/H_c$, with $H_c = 2k_1 J/g\mu_B$,
and added a constant to $\Ecl$ so that
it vanishes at the minima $(\tta,\phi) = (\tta_0,0)$ and $(\tta_0,\pi)$.
Thus, along the instanton, $\Ecl = 0$. Writing $\cos\tta_0 = u_0$, the solution
of this equation gives
\beq
\cos\tta = {u_0 +
              i \lam^{1/2} \sin\phi (1 - u_0^2 - \lam \sin^2\phi)^{1/2}
                          \by
             1 - \lam \sin^2\phi}.
\eeq

It is clear from symmetry that there are two instanton paths, which wind
about $\xhat$ in opposite directions (see Fig.~1 again).
We take $\phi(-\infty) = 0$ and 
$\phi(\infty) = \pm\pi$ for these paths. If we denote the two paths by
A and B, then the real parts of their actions ($=iJ\are$) are equal:
\bea
S_R &=& {\rm Re}S_{A,B} = J \lam^{1/2} \int_0^{\pm \pi}
              {\sin\phi (1 - u_0^2 - \lam \sin^2\phi)^{1/2}
                          \by
             1 - \lam \sin^2\phi} d\phi, \nnu\\
    &=&  J \left[
           \ln \lp {\rta + \rtl \over \rta - \rtl} \rp
           - {u_0 \over \rtlb}
              \ln \lp {\rtd + u_0 \rtl \over \rtd - u_0 \rtl} \rp
            \right].  \label{Sr}
\eea
The imaginary parts, on the other hand are necessarily unequal, since by
the interpretation of $\are$ as an area,
\beq
S_B - S_A = i J \times \Om,
\eeq
where $\Om$ [See \eno{Asurf}] is the area enclosed between A and B.
From \eno{Aline} we obtain,
\beq
\Om = \int_{-\pi}^{\pi} \left(1 - {u_0 \by 1 - \lam\sin^2\phi} \right) d\phi
       = 2\pi \left( 1 - {u_0 \by \sqrt{1-\lam}} \right).
\eeq
Since $\Dta \propto \cos(J \Om/2)$, we conclude that it vanishes whenever
\beq
{H \by H_c} = {\sqrt{1 - \lam} \by J}\Bigl[ J - n - \tshf \Bigr],
\eeq
in exact accord with \eno{plhx} (put $\ell = \ell' = 0$). Although the precise
location of the quenching points will change if the Hamiltonian is varied,
the effect is clearly general.

\subsection{Tunneling prefactors}

The above discussion does not explain how the prefactor $D$ is to be
calculated. In fact, this calculation is somewhat more subtle for spin than
it is for massive particles. If one evaluates the Gaussian fluctuations that
yield the prefactor naively, directly following the massive particle case, the
result is then not asymptotically correct as $J \to \infty$ \cite{gk92}.
This point is perhaps not of great concern for numerical estimates of
tunneling rates in a genuine physical setting, but it is nevertheless an
annoying gap in the formalism.  Although there do exist other path integral
aproaches which find the splitting correctly \cite{es86,bpp97}, the
calculations are very intricate, and the simplicity seen in the massive
particle case is lost. 

To describe these subtleties, let us first consider
the analog of the propagator (\ref{U21}) for a massive
non relativistic particle with a geometrical position coordinate $q$. This is
just the amplitude to go from a state $\ket{q_i}$ at $t=t_i$ to a state
$\ket{q_f}$ at $t=t_f$. Further, it is useful to let $q_i$ and $q_f$ be
completely general, i.e., not necessarily minima of the potential energy,
and also to consider the propagator for real time. If we write this amplitude
as a Feynman path integral \cite{rsqm}, in the semi-classical limit it is
again dominated by the classical paths, for which the action $\Scl$ is least.
Paths far away from the classical one
have phases that vary extremely rapidly under small changes of the path,
and thus these paths end up canceling each other. A great deal of quantum
mechanics can be understood just by stringing a phase factor $\exp(i\Scl/\hbar)$
on the classical trajectories. If one wants to get the actual magnitudes of
transition amplitudes, however, one must go a little further, and evaluate
the integral over the small fluctuations about the classical path. This gives
rise to the so-called {\it fluctuation\/} or {\it van Vleck determinant\/},
which is a prefactor
multiplying the phase factor $\exp(i\Scl/\hbar)$, c.f. Ref.\cite{bm72}.
(The phase and prefactor correspond approximately to the eikonal and
transport equations in the standard WKB method.)

For spin path integrals, the fluctuation determinant is more difficult,
because in contrast to the particle states for which $\tran{q}{q'} = 0$
if $q \ne q'$, spin coherent states are not orthogonal:
$\tran{\nhat}{\nhat'} \ne 0$ in general even if $\nhat \ne \nhat'$. At first
sight, it seems that one should include discontinuous paths among the
fluctuations. It turns out however, that this is not so, and by
discrete time dissection of the path integral, the analog of the van Vleck
determinant can be found provided one pays careful attention to the boundary
conditions on the spin path. This has been done by several authors%
\cite{hgs87,eak95,vvps95,spg00}, but the results do not seem to be widely
known. It seems to this author, that once the van Vleck
prefactor is understood, tunneling prefactors should be calculable with
the same degree of ease as for massive particles \cite{sc77+}, but except for
some work in Ref.~\cite{vvps95} this does not seem to have been widely
appreciated yet.

\section{The Discrete Phase Integral Method}
\label{dpi}
\subsection{Basic formalism: local Bloch waves}

The basic idea of the DPI method is to try and solve Schr\"odinger's
equation in the $J_z$ basis as a recursion relation or difference equation.
For constant coefficients, linear difference and differential equations can
both be solved in terms of exponentials.  For slowly varying coefficients, the
WKB approach is a powerful one for differential equations, and many of the
ideas used there can be carried over to difference equations. With this in
mind, let us suppose $\ket{\psi}$ is an eigenstate of $\ham$ with energy $E$.
Then, with
$J_z \ket m = m\ket m$, $\tran{m}{\psi} = C_m$,
$\mel{m}{\ham}{m} = w_m$, and $\mel{m}{\ham}{m'} = t_{m,m'}$ ($m \ne m'$),
we have
\beq
\mathop{{\sum}'}_n t_{m,n} C_n + w_m C_m = E C_m, \label{Seq}
\eeq
where the prime on the sum indicates that the term $n=m$ is to
be omitted.

We can think of Eq.~(\ref{Seq}) as a tight binding model for an
electron in a one-dimensional lattice with sites labelled by $m$,
and slowly varying on-site
energies ($w_m$), nearest-neighbor ($t_{m,m\pm 1}$) hopping terms, 
next-nearest-neighbor ($t_{m,m\pm 2}$) hopping terms, and so on.
In most interesting problems, hopping to very distant neighbors
is negligible,
and the recursion relation involves only a handful of terms. Since we can
think of dynamics in this model in terms of wavepackets, it is clear
that there is a generalization of the usual continuum quasiclassical
or phase integral method to the lattice case. This is the DPI method.

More specifically,
the DPI method is applicable to a recursion relation such as
(\ref{Seq}) whenever the $w_m$ and $t_{m,m\pm \al}$ vary sufficiently
slowly with $m$.
If these quantities were independent of $m$, the solutions to \eno{Seq}
would be Bloch waves $C_m = \exp(iqm)$, with an energy
\beq
E = w_m + 2 t_{m,m+1} \cos q + 2 t_{m,m+2} \cos 2q
    \equiv E(q), \label{Evsq}
\eeq
To save writing, it is sometimes useful to identify
$w_m \equiv t_{m,m}$, and we shall
use the notations $w_m$ and $t_{m,m}$ interchangably. If for fixed $\al$,
the $t_{m,m+\al}$ vary slowly with $m$ (where the meaning of this term
remains to be made precise), we expect it to be a good approximation to
introduce a local Bloch wavevector, $q(m)$, and write $C_m$ as an
exponential $e^{i\Phi}$, whose phase $\Phi$ accumulates approximately as
the integral of $q(m)$ with increasing $m$, in exactly the same way that
in the continuum quasiclassical method in one dimension, one writes
the wavefunction as $\exp(iS(x)/\hbar)$, and approximates $S(x)$ as the
integral of the local momentum $p(x)$.

Previous work with the DPI method \cite{dm67,sg75,pb78,pb93,vhs86}
has been limited to the case where the recursion relation has only three
terms, i.e.,  only nearest neighbor hopping is present.  As discussed by
Braun, the DPI approximation has been
employed in many problems in quantum mechanics where the Schr\"odinger
equation turns into a three-term recursion relation in a suitable
basis. All the types of problems as in the continuum case
can then be treated---Bohr-Sommerfeld quantization, barrier
penetration, tunneling in symmetric double wells, etc. In addition, one
can also use the method to give asymptotic solutions for various
recursion relations of mathematical physics, such as those for the
Mathieu equation, Hermite polynomials, Bessel functions, and so on.
The general procedures are well known and simple to state. For any
$E$, one solves the Hamilton-Jacobi and transport equations to
obtain $q(m)$ and $v(m)$, and writes $C_m$ as a linear
combinations of the independent solutions that result. The interesting
features all arise from a single fact --- that the DPI approximation
breaks down at the so-called {\it turning points}. These are points
where $v(m)$ vanishes. One must relate the DPI solutions on opposite
sides of the turning point by connection formulas, and the solution
of all the various types of problems mentioned above depends on
judicious use of these formulas.

In the \Fe8 problem, the recursion relation involves five terms.
The diagonal terms ($w_m$) arise
from the $J_z^2$ and $J_zH_z$ parts of $\ham$, the
$t_{m,m \pm 1}$ terms from the $J_xH_x$ part, and the
$t_{m, m\pm 2}$ terms from the $J_x^2$ part.
This gives rise to new features over and above the three term case.
In particular, we encounter
nonclassical turning points, i.e., turning points at $m$ values other
than those at the limits of the classically allowed motion. It is these
turning points that give rise to oscillatory tunnel splittings, so that
this effect is absent in systems described by three-term recursion
relations.

The fundamental requirement for a quasiclassical
approach to work is that $w_m$
and $t_{m,m\pm \al}$ ($\al = 1,2$) vary slowly enough with $m$ that
we can find smooth continuum approximants $w(m)$ and $t_{\al}(m)$,
such that whenever $m$ is an eigenvalue of $J_z$, we have
\bea
w(m) &=& w_m, \label{wcont} \\
t_{\al}(m) &=& (t_{m,m+\al} + t_{m,m-\al})/2, \quad \al=1,2.
   \label{tcont}
\eea
We further demand that
\beq
{dw \over dm} = O\left( {w(m) \over J } \rp, \quad
{dt_\al \over dm} = O\left( {t_{\al}(m) \over J } \rp,
\label{slow}
\eeq
with $m/J$ being treated as quantity of order 1. Problems for which these
conditions cannot be met are not amenable to the DPI method. It is not
difficult to see that for Eqs.~(\ref{ham}) and (\ref{ham4}),
these conditions will hold in the semiclassical limit $J \gg 1$.

Given these conditions,
the basic approximation, which readers will
recognize from the continuum case, is to write the
wavefunction as a linear combination of the quasiclassical forms
\beq
C_m \sim {1 \over \sqrt{v(m)}}\exp\lp i\int^m q(m') dm'\rp,
    \label{Cwkb}
\eeq
where $q(m)$ and $v(m)$ obey the equations
\bea
E &=& w(m) + 2t_1(m) \cos q + 2t_2(m) \cos(2q)
     \equiv \hsc(q,m), \label{hjeq} \\
v(m) &=& \ptl \hsc/\ptl q = -2\sin q(m)
           \big(t_1(m) + 4 t_2(m) \cos q(m)\bigr).
     \label{vm}
\eea
Equations (\ref{hjeq}) and (\ref{vm}) are the lattice analogs of the
eikonal and transport equations. Equation (\ref{Cwkb}) represents
the first two terms in an expansion of $\log C_m$ in powers of
$1/J$.

\subsection{Turning points and connection formulas}

The basic DPI approximation fails whenever
$v(m) = \ptl\hsc(q,m)/\ptl q = 0$, because then it diverges.
We shall call all such points
{\it turning points} in analogy with the continuum case. In contrast
to that case, however, we will find that turning points
are not just the limits of the classical motion for a given energy,
once the notion of the classically accessible region is suitably
understood.

Since we must also obey the eikonal equation (\ref{hjeq}) in addition
to the condition $v(m) = 0$, at a turning point both $m$ and $q$ are
determined if $E$ is given.  Setting $v=0$ in \eno{vm}, we see that we
must have either $q=0$, or $q=\pi$, or $q = q_*(m)$, where
\beq
\cos q_*(m) = - t_1(m)/4t_2(m). \label{cosqstar}
\eeq
Substituting these values of $q$ in the eikonal equation, we see that
a turning point arises whenever
\beq
E = U_0(m),\ U_{\pi}(m),\ {\rm or}\ U_*(m),
   \label{Econd}
\eeq
where,
\bea
U_0(m) &=& \hsc(0,m) = w(m) + 2t_1(m) + 2t_2(m), \label{U0} \\
U_{\pi}(m) &=& \hsc(\pi, m) = w(m) - 2t_1(m) + 2t_2(m), \label{Upi} \\
U_*(m) &=& \hsc(q_*,m) = w(m) - 2t_2(m) - {t_1^2(m) \over 4 t_2(m)}.
           \label{Ustar}
\eea
Note that $q_*(m)$ may be complex for some $m$, but since
$\cos q_*$ is always real, $U_*$ is real for all $m$. We shall
refer to these three energy curves as {\it critical curves}.
We shall see that they collectively
play the same role as the potential energy in the continuum
quasiclassical method.

To better understand the turning points, let us assume
that $t_1 < 0$, and $t_2 > 0$. [This is the case for the Hamiltonian
(\ref{ham}). We can always arrange for $t_1$ to be negative by means
of the gauge transformation $C_m \to (-1)^m C_m$. Thus there is
only one other case to be considered, namely, $t_1<0$, $t_2<0$.
This is discussed in \rno{agjmp00}.] It then follows that
$U_{\pi} > U_0$, and that
\beq
U_0(m) - U_*(m) = {1 \over 4t_2(m)}
        \bigl(t_1(m) + 4t_2(m) \bigr)^2 \ge 0. \label{difU}
\eeq
Secondly, let us think of
$\hsc(q,m)$ for fixed $m$ as an energy band curve. Then $U_{\pi}$ is
always the upper band edge, while the lower band edge is either
$U_0$ or $U_*$ according as whether $-t_1/4t_2$ is greater than or
lesser than 1. To deal with this possibility, it pays to introduce
a dual labeling scheme for all three curves $U_0$,
$U_{\pi}$, and $U_*$. We write $U_{\pi}(m) \equiv U_+(m)$,
and
\bea
U_0(m) = U_i(m),\  U_*(m) = U_-(m),\quad{\rm if}\ q_* \in (0,\pi),
                                             \label{Ui} \\
U_0(m) = U_-(m),\  U_*(m) = U_f(m),\quad{\rm if}\ q_* \not\in (0,\pi).
                                             \label{Uf}
\eea
The subscripts $+$ and $-$ denote upper and lower band edges, while
the subscripts {\it i} and {\it f} denote {\it internal} and
{\it forbidden} respectively, since in the first case above, $U_0$
lies inside the%
\begin{figure}
\centerline{\psfig{figure=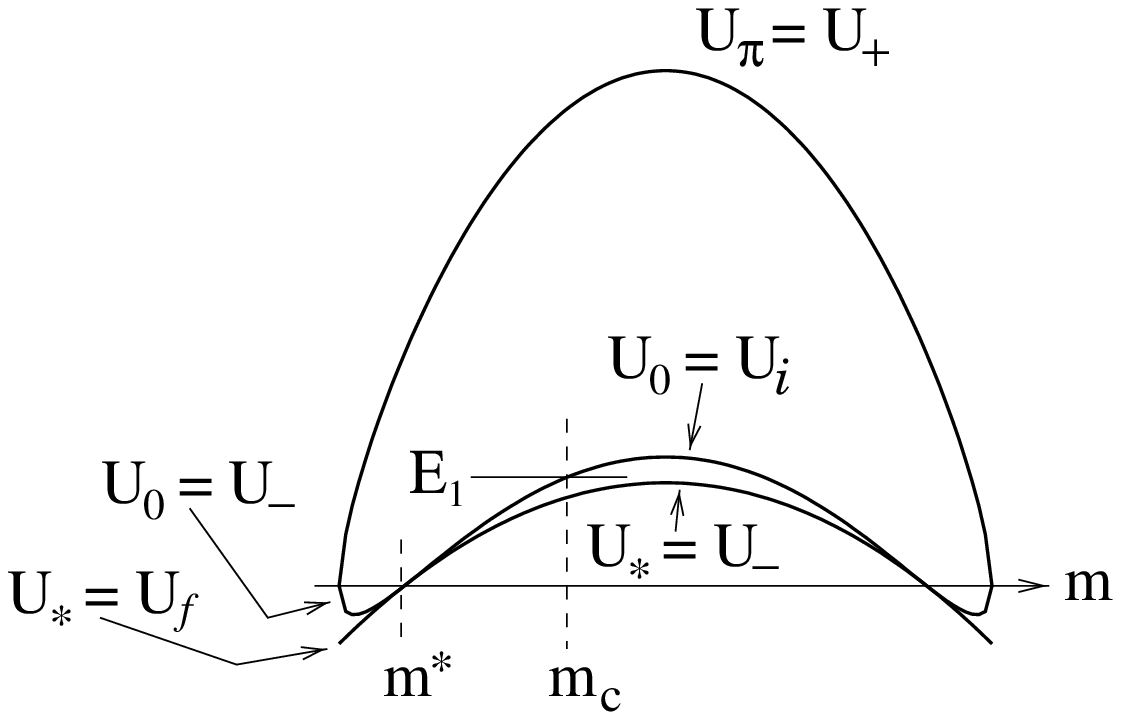,height=5cm,width=0.44\linewidth}
\psfig{figure=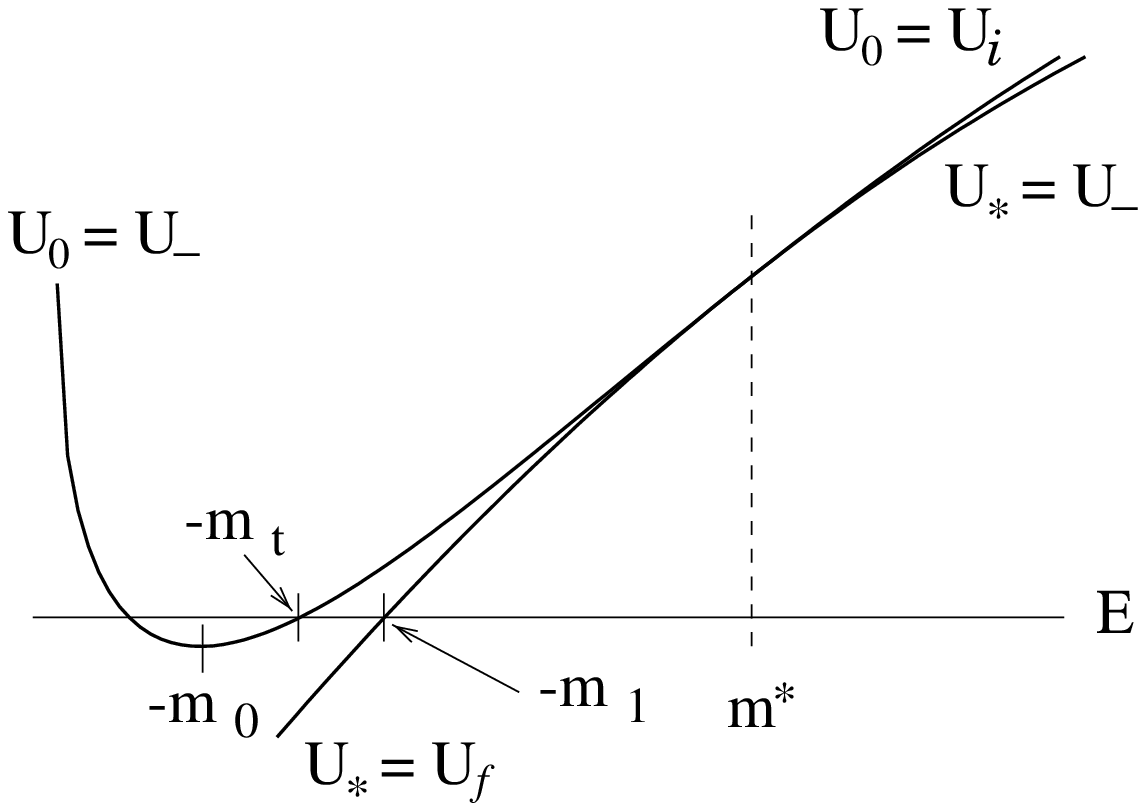,height=5cm,width=0.48\linewidth}}
\caption{Critical energy curves for the \Fe8 Hamiltonian when
$\bH\|\xhat$, showing the dual labeling scheme. In the right hand
figure, the region near the left minimum of $U_0$ is magnified,
showing the two turning points $-m_t$ and $-m_1$.}
\end{figure}
\noi energy band, while in the second case, $U_*$ lies outside.
As examples of these curves for a symmetric recursion relation,
we show those for \Fe8 in Fig.~10, along with a magnified view of the
lower left hand portion.

Turning points where $E=U_+$, or $E=U_-$ when $U_- = U_0$, are
analogous to those encountered in the continuum quasiclassical
method, since the energy lies at a limit of the classically allowed
range for the value of $m$ in question. Points where $E=U_-$ when
$U_- = U_*$ are physically analogous, but mathematically different
since the value of $q_c$ is neither $0$ nor $\pi$. Points where
$E = U_i$ (see the energy $E_1$ in Fig.~10, e.g.) are novel in that
the energy is {\it inside} the
classically allowed range for $m_c$, but the mathematical form
of the connection formulas is identical to the case $E=U_-=U_0$
since $q_c = 0$. Most interesting are the turning points with
$E = U_f$ (the point $m=-m_1$ in Fig.~10, for instance),
since now the energy is outside the allowed range
for $m=m_c$, and the value of $q_c$ is therefore necessarily complex.
These points lie ``under the barrier" and turn out to be the ones of
importance for understanding oscillatory tunnel splittings.
The derivation of connection formulas at these turning points is not
particularly difficult, but quite lengthy. In fact, it is quite lengthy
to even quote the connection formulas themselves, and we refer readers
to \rno{agjmp00} for the details. The key point is that a solution with
a purely imaginary $q$ on side of the turning point, representing, let us
say a solution that decays with growing $m$, turns on the other side into
a solution with a {\it complex\/} $q$ with a nonzero real part, representing
a solution that decays with an oscillating envelope as $m$ increases. These
kinds of wavefunctions can simply never arise in conventional one-dimensional
continuum problems.

From this point on, the exercise of calculating tunnel splittings is
exactly as in the continuum case. One demands that the wavefunctions
decay as $m \to \pm\infty$ and applies connection formulas at the turning
points to continue the wavefunction into the interior. Finally near $m=0$,
one demands that the wavefunctions so found from the two sides should
agree. This gives the eigenvalue condition. The one new complication (besides
the new turning points) is that at each $m$, there are {\it four\/} solutions
to the Hamilton-Jacobi equation instead of two in the continuum case. As
$m \to -\infty$, all four $q$'s are pure imaginary, but of these only the
two which are negative imaginary can be kept. Thus at every point $m$, one
must keep track of two DPI solutions. An exception to this arises in the
case of symmetric double wells ($\bH \| \xhat$), where Herring's formula
\cite{ch62}
simplifies matters, and one can work purely with a wavefunction that decays
away from one of the wells in {\it both\/} directions. This formula is
sufficiently important to be worth discussing separately. We do this in the
next subsection, present results that follow from its application in the
succeeding subsection, and for the asymmetric case in the one following that.

\subsection{The symmetric case; Herring's formula}

It is useful to recall a basic formula for tunnel splittings in symmetric
potentials [$V(-x) = V(x)$] for massive particles \cite{llqm2}. Let the 
minima be at $x = \pm a$, and let
$\psi_0(x)$ be a wavefunction localized entirely in the right hand well,
with an energy $E_0$. This wavefunction obviously does not obey 
Schrodinger's equation near the left hand well, and we could imagine modifying
the potential suitably in that region so that the wavefunction continues to
decay wtih decreasing $x$ even in the vicinity of the left well. Since
we will never need $\psi_0(-a)$, the precise way in which this is done is
not important.

If $E_0$ is well below the barrier, then we expect there to be two states
$\psi_s$ and $\psi_a$, with energies $E_1$ and $E_2$, both very close to
$E_0$, and with wavefunctions that are very accurately given by
\beq
\psi_{s,a} = {1 \by \sqrt2}\bigl[ \psi_0(x) \pm \psi_0(-x) \bigr].
\eeq
The product $\psi_0(x)\psi_0(-x)$ is everywhere exponentially small, and
so therefore, if the wavefunction $\psi_0(x)$ is normalized, so are
$\psi_{s,a}$. For the same reason,
\beq
\int_0^{\infty} \psi_0(x) \psi_a(x)\, dx = {1 \by\sqrt2}.
\eeq

The differential equations obeyed by $\psi_0$ and $\psi_a$ in the region
$x > 0$ are
\bea
-(\hbar^2/2m) \psi''_0(x) + V(x) \psi_0(x) &=& E_0 \psi_0(x), \\
-(\hbar^2/2m) \psi''_a(x) + V(x) \psi_a(x) &=& E_2 \psi_a(x).
\eea
We multiply the first equation by $\psi_a$, the second by $\psi_0$, 
subtract, and integrate from $0$ to $\infty$. This yields
\bea
E_2 - E_0 &=& \sqrt2 {\hbar^2 \by 2m}
                  \int_0^{\infty}(\psi_a\psi_0'' - \psi_0\psi_a'')dx \nnu\\
                   &=& \sqrt2 {\hbar^2 \by 2m}
                  [\psi_a\psi_0' - \psi_0\psi_a']_0^{\infty} \nnu \\
                   &=& {\hbar^2 \by m} \psi_0(0) \psi'_0(0),
\eea
where we have used the facts that $\psi_{0,a}(\infty) = 0$, $\psi_a(0) =0$,
and $\psi'_a(0) = {\sqrt 2}\psi'_0(0)$. A similar calculation yields
$E_1 - E_0 = - (E_2 - E_0)$, so that the energy splitting, $\Dta$ is given
by
\beq
\Dta = {2\hbar^2 \by m} \psi_0(0) \psi'_0(0).
    \lbl{Herrcont}
\eeq
This is Herring's formula. It has a nice and physically appealing
interpretation in terms of the probability current at $x=0$.

If we now use the WKB method to find $\psi_0(0)$ and $\psi'_0(0)$, keeping
in mind the normalization, for the $n$th pair of levels, we get
\beq
\Dta_n = g_n {\hbar\om \over \pi}
              \exp\left[ - \int_{-a'_n}^{a'_n} {|p|\over \hbar}dx
                         \right]. \label{Dtan}
\eeq
Here, $\om$ is the small oscillation frequency in the well,
$[V''(a)/m]^{1/2}$, $\pm a'_n$ are the classical turning points for the $n$th
energy level pair with the mean value $E_0 \apx (n+\hf)\hbar\om$, and
\beq
g_n = {\sqrt{2\pi} \over n!} \lp n + \tshf \rp^{n +\hf}
         e^{-(n + \hf)}, \label{gndef}
\eeq
Note that if $g_n$ were unity, we would have the formula of
Ref.~\cite{llqm2}. The corrections are small:
$g_0 = (\pi/e)^{1/2} \approx 1.075$, $g_1 \approx 1.028$,
$g_2 \approx 1.017$, and so on. Stirling's formula for $n!$
shows that $g_n \to 1$ as $n \to \infty$. These corrections are well known
to many workers, but not known to many others, and they arise from the fact
that as $n$ becomes smaller, the use of a linear turning point formula at
the turning point is increasingly inaccurate. The correct procedure is
to use quadratic turning point formulas \cite{bm72}. A pedagogical
discussion of this matter may be found in Ref.~\cite{agajp00}.

One can repeat the above argument for the discrete case making the obvious
modifications as needed. The analogue of \eno{Herrcont}, Herring's formula, is
\beq
\Dta = \cases{
        2\left[ t_{0,1} C_0(C_1 - C_{-1}) + t_{0,2}C_0(C_2 - C_{-2})
                + t_{-1,1} (C_1^2 - C_{-1}^2) \right],
                                & \\integer $J$, \cr
        2\ t_{-\hf,\hf} \Bigl( C^2_{\hf} - C^2_{-\hf}\Bigr)
         + 4\ t_{-\tbt,\hf} \lp C_{\hf}C_{\tbt} - C_{-\hf}C_{-\tbt}\rp,
                                & half-integer $J$. \cr}
 \label{Herr}
\eeq
When it is combined with the DPI approximation, one obtains for both
integer and half-integer $J$ \cite{agprb00_1},
\beq
\Dta = {\om_0 g_n \over 2\pi}
      \left[ \exp \lp i\int_{-m_t}^{m_t} q(m') dm' \rp
               + \ \comp \right].
     \label{Dgen3}
\eeq
The similarity to the continuous case is striking. The only point worth
remarking is the possibility of having the Bloch vector $q(m)$ be complex
in part or all of the tunneling region. In \eno{Dgen3}, we have written
the formula for the \Fe8 problem, where there are two $q$'s with positive
imaginary parts and equal and opposite real parts in the tunneling region.

It may be useful at this point to make a general remark about how tunneling
prefactors are obtained in the DPI method. Equations (\ref{Dtan}) and
(\ref{Dgen3}) contain these prefactors naturally, and it is apparent that
they arise as a consequence of the $1/\sqrt{v(m)}$ normalization in the
general DPI form and the connection formulas at turning points. Provided
one can solve the transport equation and apply the connection formulas
correctly, these factors are relatively easy to obtain. (We shall see that
this continues to be true in the asymmetric case.) Historically, however,
one error which has been made is to use linear turning point formulas
all throughout, as in Ref.~\cite{llqm2}. As already mentioned, this leads
one to miss the curvature factors $g_n$. This error is widely repeated,
perhaps because of the reputation of this text, and perhaps because it is
numerically insignificant. From a logical point of view, however, there
is no reason not to incorporate the curvature corrections, and they
{\it must\/} be kept if one wants an answer for the tunneling amplitude
itself (and not just its logarithm) that is properly asymptotic
to the true answer as $\hbar \to 0$, or as $J \to \infty$.

\subsection{Results for \Fe8 in symmetric case [26]}

To apply the general results to \Fe8, we need the explicit forms for
$w(m)$, and $t_{\al}(m)$. It is convenient  to measure energies
(including $\om_0$) in units of $k_1\baj^2$, and introduce the scaled
variable $\mu = m/\baj$, where
\beq
\baj = J +\tshf.
\eeq
In these variables, when $\bH\|\xhat$, we get
(with $h_x = JH_x/\baj H_c$)
\bea
w(m) &=& (1+\lam)(1- \mu^2)/2, \label{wofm} \\
t_1(m) &=& -h_x (1-\mu^2)^{1/2}, \label{t1ofm} \\
t_2(m) &=& (1-\lam)(1 - \mu^2)/4. \label{t2ofm}
\eea

The actual evaluation of the integrals is mostly a matter of careful but
straightforward analysis, and the final answer for the splitting can be
written as
\beq
\Dta_n = {1\over n!} \sqrt{8\over\pi} \om_0 F^{n+\hf}
         e^{-\Gam_0}\cos\Lam_n,
      \label{Dnfinal}
\eeq
where,
\bea
\om_0 &=& 2J[k_1k_2(1-h_{x0}^2)]^{1/2}, \label{om0fe8}\\
F &=& 8 J {\lam^{1/2} (1 - h_x^2)^{3/2} \over 1 - \lam - h_x^2},
        \label{Ffe8} \\
\Gam_0 &=& \baj \left[
           \ln \lp {\rta + \rtl \over \rta - \rtl} \rp
           - {h_x \over \rtlb}
              \ln \lp {\rtd + h_x \rtl \over \rtd - h_x \rtl} \rp
            \right],  \label{Sr2} \\
\Lam_n &=& {{\rm max}} \left\{0,\
                 \pi J\lp 1 - {H_x \over \sqrt{1-\lam}H_c} \rp
              - n\pi \right\}.
   \label{Lamtot}
\eea
In Eqs.~(\ref{om0fe8}--{\ref{Lamtot}), $\lam = k_2/k_1$, and
\beq
h_x = {JH_x \over \baj H_c}, \quad h_{x0} = {H_x \over H_c}.
         \label{hxhx0}
\eeq
Recall that $H_c = 2k_1 J/g\mu_B$.

It is an immediate consequence of these results that $\Dta_n$ vanishes
when
\beq
{H_x \over H_c} = {\sqrt{1-\lam} \over J}
                  \left[J - \ell - \hf \right], \label{quench2}
\eeq
with $\ell = n$, $n+1$, $\ldots$, $2J - n -1$. These are precisely the
results quoted in Sec.~II.

\subsection{Results for \Fe8 in asymmetric case [27]}

When $\bH$ has components both along $\xhat$ and $\zhat$, the critical
curves are no longer symmetric (Fig.~11), and the problem cannot
be reduced to just quoting a splitting. Suppose we consider an energy $E$,
as drawn in Fig.~11, and suppose that it leads
to turning points at $m'_a$, $m'_b$, $m'_c$ on the left hand
side, and $m''_c < m''_b < m''_a$ on the right. (We 
denote quantities pertaining to the left hand solution or the left
hand side of the well by either a single prime or a suffix%
\begin{figure}
\centerline{\psfig{figure=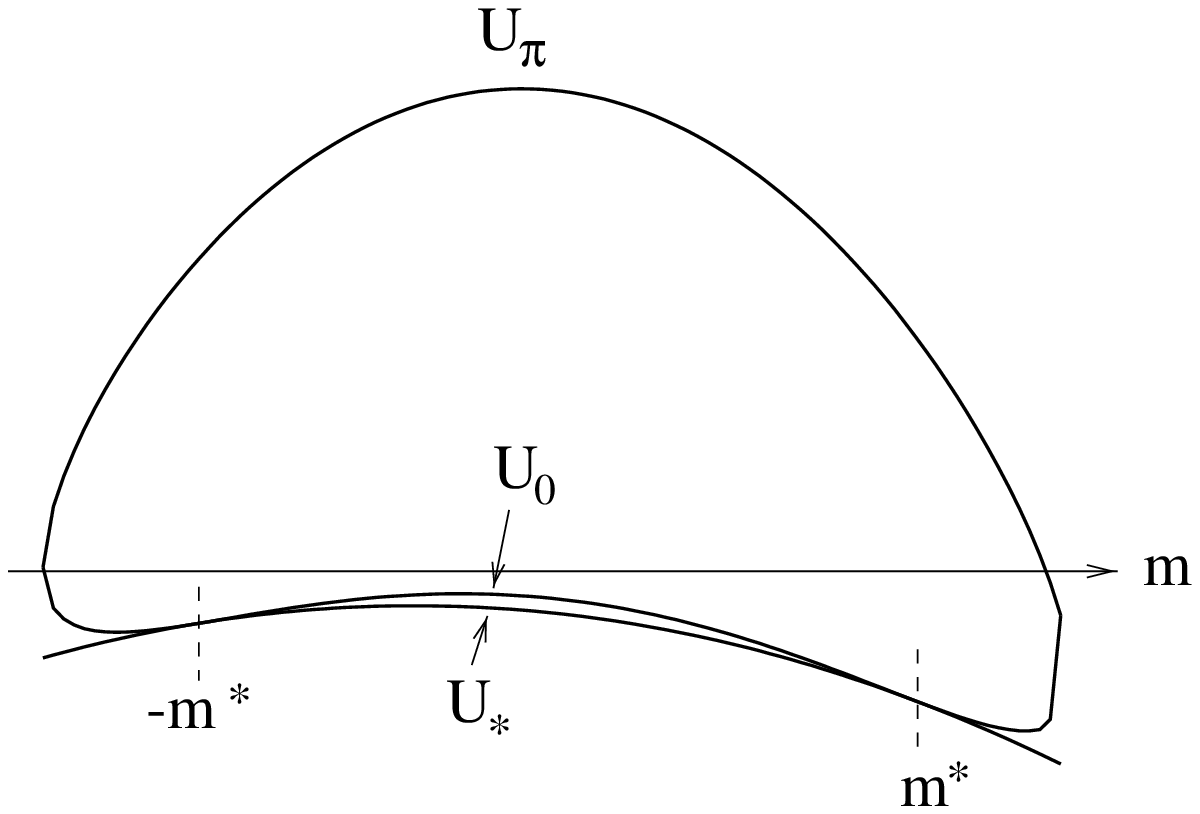,height=5cm,width=0.46\linewidth}
\psfig{figure=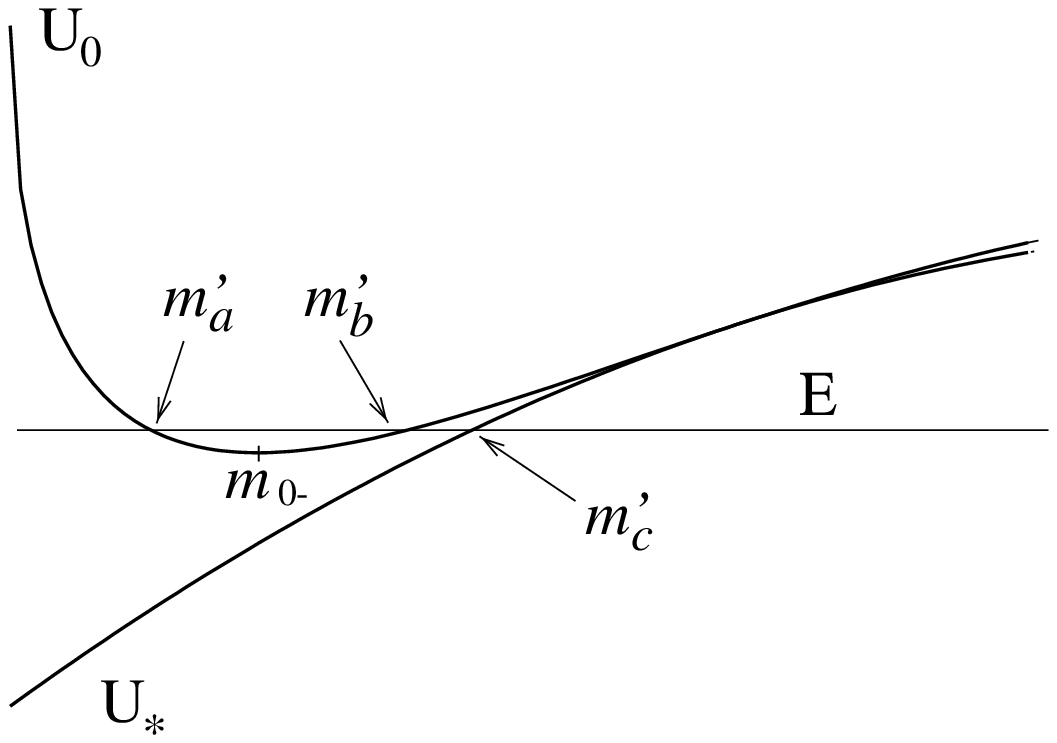,height=5cm,width=0.46\linewidth}}
\caption{Critical energy curves for the \Fe8 Hamiltonian when
$\bH$ has both $x$ and $z$ components.  In the right hand
figure, the region near the left minimum of $U_0$ is magnified,
showing the various turning points.}
\end{figure}
\noi $-$, and corresponding right-hand-well quantities by a double
prime or a $+$
suffix.) A wavefunction $C'_m$ which decays as $m \to -\infty$ will possess
the following additional characteristics. It will either oscillate or
continue to have the same exponential character in the classically allowed
region $m'_a < m < m'_b$. In the region just to the
right of $m'_b$ it will consist of a decaying part and a growing part.
The new feature will be encountered at $m'_c$ where $E = U_*$. For
$m > m'_c$, both the growing and decaying parts will acquire oscillatory
envelopes. Similar remarks apply to the right side wavefunction $C''_m$.
The eigenvalue condition will be obtained by matching the wavefunction in
the central region $m'_c \ll m \ll m''_c$.

Let us now denote the left well minimum by $m_{0-}$, the small oscillation
frequency in that well by $\om_{0-}$, and the value of $U_0(m_{0-})$
by $E_-$. We write
\beq
E = E_- + (\nu' + \tshf) \omzm. \label{defnu'}
\eeq
In the same way we define analogous quantities for the right hand well.
We expect that tunneling will be significant only when the energy $E$ is
such that it matches a level in both wells, say, level numbers $\ell'$
and $\ell''$, where $\ell=0$ is the ground level. Since the problem is
harmonic near the two potential wells, we can write the energy levels
using a harmonic oscillator formula, and with this in mind, we define
an offset $\eps'$,
\beq
\nu' = \ell' + {\eps'\over \omzm}, \label{defeps}
\eeq
where $\ell'$ is a positive integer, and $\eps'$ 
lies in the interval $(-1/2, 1/2)\omzm$. $\eps''$ is similarly defined.
Our remarks above mean that mixing between the two wells will be significant
only when $\eps'$ and $\eps''$ are both small. This is indeed the case, and
the matching or eigenvalue condition turns out to be
\beq
\eps'\eps'' = {1\by 4} [\Dll]^2, \label{eval}
\eeq
where $\Dll$ is an exponentially small tunneling amplitude that we shall 
give shortly. The essential point emerges if we define
\bea
\eps &=& \tshf(\eps'+\eps'')
          = E - \hf\lp E_- + E_+ + (\ell'+\tshf)\omzm
                     +(\ell''+\tshf)\omzp \rp, \label{Eavg} \\
\dta &=& \eps'' - \eps'
          = \lp E_- - E_+ + (\ell'+\tshf)\omzm
                     - (\ell''+\tshf)\omzp \rp. \label{Edif}
\eea
Then, the eigenvalue condition \eno{eval} can be rewritten as
\beq
\eps = \pm \tshf [\dta^2 + \Dta^2(\ell',\ell'') ]^{1/2}. \label{diab}
\eeq
These are of course, the eigenvalues of the two-level-system Hamiltonian
\beq
\ham_{\rm TLS} = \hf \lp
                 \begin{array}{cc}
                  \dta & \Dll \\
                  \Dll & -\dta   \end{array} \rp, \label{htls}
\eeq
exactly as we should expect. 
The quantity $\eps$ is the energy measured from a convenient
reference point, while $\dta$, which depends on the fields $H_x$,
$H_z$, and the quantum numbers $\ell'$ and $\ell''$ of the states
whose mixing is being examined, is the bias or offset between these energy
levels in the absence of tunneling. Equation (\ref{defDta}) below
gives the tunneling amplitude between these levels when the bias
is small, i.e., when the two levels are in approximate degeneracy.
Note that although this amplitude is defined even for relatively
large biases---biases comparable to the intrawell spacings
$\ompm$---and indeed is not very sensitive to the value of the
bias, the concept of tunneling is physically sensible and useful
only when the bias is comparable to or less than the amplitude
$\Dta$. If $\dta \gg \Dta$, we get $\eps \apx \pm \dta/2$, i.e.,
$\eps'' \apx \dta$, $\eps' \apx \Dta^2/\dta$,
or the other way around. Choosing the first way, we find that
$E\apx E_- + (\ell'+\tshf)\omzm$, and a wavefunction essentially
given by $(1\quad \Dta^2/\dta)^{\rm T}$,
i.e., localized in the left well, with negligible mixing with the right well.

It remains to give the expression for $\Dll$. This is,
\beq
\Dll = {2 \over \pi} (g_{\ell'}g_{\ell''})^{1/2}
                      (\omzm\omzp)^{1/2} e^{-\Gam_G}\cos\Lam_c,
    \label{defDta}
\eeq
with $\Gam_G$ being the total Gamow factor
\beq
\Gam_G = \int_{m'_b}^{m''_b} |\,{\rm Im}\,q_1(m)|\,dm, \label{Gamow}
\eeq
and $\Lam_c$ being the phase integral
\beq
\Lam_c = \int_{m'_c}^{m''_c} |\,{\rm Re}\,q(m)|\, dm.
  \label{Lamc}
\eeq
In Eq.~(\ref{Gamow}) the subscript 1 on $q(m)$ means that we are to take
that solution of the Hamilton-Jacobi equation which goes to zero at
$m'_b$ and $m''_b$. In \eno{Lamc}, on the other hand, we can choose
any of the four solutions for the wavevector, since they are of the form
\beq
q(m) = \pm i\kap(m) \pm \chi(m),
\eeq
where $\kap(m)$ and $\chi(m)$ are both real, and the signs can be chosen
independently from the two $\pm$ options.

The application of these results to \Fe8 again requires doing a certain
number of integrals. The problem of greatest
interest is the location of the diabolical points, and for that we need
only solve the conditions $\dta= \Dta = 0$. The latter condition can
only come about when $\Lam_c$ is an odd multiple of $\pi/2$, so in fact
we need only give formulas for $\dta$ and $\Lam_c$. We find
\beq
\dta(h_z,\ell',\ell'') = 4 \mu_0 h_z
        + {2 \sqrt{\lam} \mu_0 \over \baj} (\ell' - \ell'')
        -{\sqrt{\lam} h_z \over \baj}(\ell'+\ell'' + 1)c_1(h_x)
        + O(\baj^{-3}), \label{ansdta}
\eeq
where
\beq
c_1(h_x) = {1 - h_x^2 + \lam(1 + 2h_x^2)
             \over \lam (1-h_x^2)}, \label{defc1}
\eeq
with $h_z = JH_z/\baj H_c$, and $\mu_0 = (1 - h_x^2)^{1/2}$. For $\Lam_c$,
we have up to order $J^0$,
\beq
\Lam_c = {\pi \over 2}\left[
             2J - (\ell' + \ell'') -
               2J{H_x \over H_c \sqrt{1-\lam}} \right]. \label{ansLam}
\eeq
If we ignore the correction $c_1(h_x)$, the conditions $\dta = \Dta = 0$
are thus precisely those quoted in Sec.~II.

\acknowledgments
I am grateful to W.~Wernsdorfer for supplying me with Figs.~2 and 4.
This work is supported by the NSF via grant number DMR-9616749.


\begin{references}
\bibitem[*]{byline} Electronic address: agarg@nwu.edu

\bibitem{sc98} S. Creagh, {\it Tunneling in Two Dimensions\/},
    in {\it Tunneling in Complex Systems\/}, edited
    by Steven Tomsovic (World Scientific, Singapore, 1978).

\bibitem{mw86} \citn{M.~Wilkinson}{Physica}{21 D}{341}{1986}.

\bibitem{ks78} I.~Ya. Korenbilt and E.~F. Shender, Zh. Eksp. Teor. Fiz.
75, 1862 (1978) [Sov. Phys. JETP 48 (1978)].

\bibitem{cg88} \citn{E.~M. Chudnovsky and L. Gunther}{Phys. Rev. Lett.}
{60}{661}{1988}.

\bibitem{bc90} B.~ Barbara and E.~M. Chudnovsky, Phys. Lett. A 145, 205 (1990).

\bibitem{gb95} Leon Gunther and Bernard Barbara, Eds., {\it Quantum Tunneling of
   Magnetization -- QTM '94}, NATO ASI Series E: Applied Sciences, Vol. 301,
   Proceedings of the NATO Advanced Research Workshop on Quantum Tunneling of
   Magnetization -- QTM '94, Grenoble and Chichilianne, France, June 27 - July
   2, 1994 (Kluwer Academic, Dordecht, 1995).

\bibitem{ns94} M.~A. Novak and R.~Sessoli, in Ref.~\cite{gb95}.

\bibitem{fs96} J. Friedman, M. P. Sarachik, J. Tejada, and R. Ziolo, Phys. Rev.
Lett. {\bf 76}, 3830 (1996).

\bibitem{tlbd96} L. Thomas et al., Nature {\bf 383}, 145 (1996).

\bibitem{vhsr94}J. Villain, F. Hartman-Boutron, R. Sessoli, and
A. Rettori, Europhys. Lett.  {\bf 27}, 159 (1994).

\bibitem{agprl98} A.~Garg, Phys.\ Rev.\ Lett.\ {\bf 81}, 1513 (1998).

\bibitem{ws99} \citn{W.~Wernsdorfer and R.~Sessoli}{Science}{284}{133}{1999}

\bibitem{ldg92} \citn{D.~Loss, D.~P. DiVincenzo, and G.~Grinstein}{\prl}
                 {69}{3232}{1992}.

\bibitem{vdh92} J.~von Delft and C.~L. Henley, \prl {\bf 69}, 3236 (1992).

\bibitem{agepl93} A.~Garg, Europhys.\ Lett.\ {\bf 22}, 205 (1993).

\bibitem{jrk78} J.~R.~Klauder, in {\it Path Integrals\/} (Proceedings of
   the NATO Advanced Summer Institute), edited by G.~J.~Papadopoulos and
   J.~T.~Devreese (Plenum, NY, 1978).

\bibitem{jrk79} J.~R.~Klauder, Phys. \ Rev.\ D {\bf 19}, 2349 (1979).

\bibitem{dm67} R.~B. Dingle and J. Morgan, Appl.\ Sci.\ Res.\ {\bf 18},
221 (1967); {\it ibid.} {\bf 18}, 238 (1967).

\bibitem{sg75} K. Schulten and R.~G. Gordon, J.\ Math.\ Phys.\ {\bf 16}, 1971 (1975).

\bibitem{pb93} P.~A. Braun, Rev.\ Mod.\ Phys.\ {\bf65}, 115 (1993).

\bibitem{vhs86} J.~L. van Hemmen and A.~S\"ut\H o,
(a) Europhys.\ Lett.\ {\bf 1}, 481 (1986);
(b) Physica {\bf 141B}, 37 (1986).

\bibitem{vhw88} J.~L. van Hemmen and W.~F. Wreszinski,
        Commun.\ Math.\ Phys.\ {\bf 119}, 213 (1988).

\bibitem{vhs95} J.~L. van Hemmen and A.~S\"ut\H o, in Ref.~\cite{gb95}.

\bibitem{agprl99} A.~Garg, Phys.\ Rev.\ Lett.\ {\bf 83}, 4385 (1999).

\bibitem{vf00} J.~Villain and Anna Fort, Euro. Phys. J. B {\bf 17}, 69 (2000).

\bibitem{agprb00_1} A.~Garg, to appear in Phys. Rev. B; cond-mat/0003114.

\bibitem{agprb00_2} A.~Garg, to appear in Phys. Rev. B; cond-mat/0003156.

\bibitem{wpj84} \citn{K.~Wieghardt, K.~Pohl, I.~Jibril, and G.~Huttner}
        {Angew. Chem. Int. Ed. Engl.}{23}{77}{1984}

\bibitem{d+93} \citn{C. Delfs et al.}{Inorg. Chem.}{32}{3099}{1993}.

\bibitem{bdg96} \citn{A.~-L. Barra,  P.~Debrunner, D.~Gatteschi, Ch.~E.
Schultz, and R.~Sessoli}{Europhys. Lett.}{35}{133}{1996}.

\bibitem{sop97} \citn{C.~Sangregorio, T.~Ohm, C.~Paulsen, R.~Sessoli,
       and D.~Gatteschi}{Phys. Rev. Lett.}{78}{4645}{1997}.

\bibitem{cam98} R.~Caciuffo, G.~Amoretti, A.~Murani, R.~Sessoli,
A.~Caneschi, and D.~Gatteschi, Phys.\ Rev.\ Lett.\ {\bf 81}, 4744 (1998).

\bibitem{osp98} \citn{T.~Ohm, C.~Sangregorio, and C.~Paulsen}{Eur. Phys. J. B}
            {6}{195}{1998}.

\bibitem{lzs32} \citn{L.~D. Landau}{Phys.\ Z. Sowjetunion}{2}{46}{1932};
   \citn{C.~Zener}{Proc.\ Roy.\ Soc. London A}{137}{696}{1932};
   \citn{E.~C. G. St\"uckelberg}{Helv.\ Phys.\ Acta}{5}{369}{1932}.

\bibitem{llqm} L.~D. Landau and E.~M. Lifshitz, {\it Quantum Mechanics}
3rd edition, (Pergamon, New York, 1977), Sec. 90. 

\bibitem{wsc00}  \citn{W.~Wernsdorfer, R.~Sessoli, A.~Caneschi,
D.~Gatteschi, and A.~Cornia}{\jap}{87}{5481}{2000}.

\bibitem{vnw29} J.~von Neumann and E.~P. Wigner, Physik.\ Z.\
{\bf 30}, 467 (1929).

\bibitem{via78} V.~I. Arnold, {\it Mathematical Methods of Classical
Mechanics} (Springer-Verlag, New York, N.Y., 1978). See Appendix 10.
More precisely, the theorem is that the {\it codimension} of
the space of a double degeneracy is 2 or 3, when the dimensionality
of the parameter space exceeds 2 for real, or 3 for complex
Hamiltonians, respectively. Another lucid and perceptive 
discussion may be found in C.~A. Mead,
J. Chem. Phys. {\bf 70}, 2276 (1979).

\bibitem{bw84} M.~V. Berry and M.~Wilkinson, Proc.\ Roy.\ Soc.\ Lond.\ A
{\bf 392}, 15 (1984).

\bibitem{hlh63} G.~Herzberg and H.~C. Longuet-Higgins,
Discuss.\ Faraday Soc.\ {\bf 35}, 77 (1963).

\bibitem{agprb95} A.~Garg, Phys.\ Rev.\ B {\bf51}, 15161 (1995).

\bibitem{kg00} Ersin Ke\c{c}ecio\u{g}lu and A.~Garg, to appear in
Phys. Rev. B; cond-mat/0003319.

\bibitem{mvb84} M.~V. Berry, Proc.\ R.\ Soc.\ Lond.\ A {\bf 392}, 45 (1984).

\bibitem{jl67} \citn{J.~S. Langer}{Ann.\ Phys. (N.Y.)}{41}{108}{1967}.

\bibitem{sc77+} \citn{S.~Coleman}{\prd}{15}{2929}{1977}; 
        {\it Aspects of Symmetry\/} (Cambridge University Press, Cambridge,
         1985), Chap. 7.

\bibitem{rsqm} R.~Shankar, {\it Principles of Quantum Mechanics\/}
(Plenum, New York, 1980), Chap. 8.

\bibitem{dh83+} F.~D. M. Haldane, Phys.\ Lett.\ A {\bf 93}, 464 (1983);
        \prl {\bf 50}, 1153 (1983); \jap {\bf 57}, 3359 (1985).

\bibitem{ef91} E.~Fradkin, {\it Field Theories of Condensed Matter Systems\/}
           (Addison-Wesley, Redwood City, 1991), Chap. 5.

\bibitem{gk92} \citn{A.~Garg and G.-H. Kim}{\prb}{45}{921}{1992}.

\bibitem{es86} M. Enz and R. Schilling, J. Phys. C 19, L711 and 1765 (1986).

\bibitem{bpp97} V.~I. Belinicher, C.~Providencia, and J.~da Providencia,
J.\ Phys.\ A: Math.\ Gen.\ {\bf30}, 5633 (1997).

\bibitem{bm72} M.~V. Berry and K.~E. Mount, Rep.\ Prog.\ Phys.\ {\bf 35},
315 (1972).

\bibitem{hgs87} \citn{H.~G. Solari}{\jmp}{27}{1097}{1987}.

\bibitem{eak95} \citn{E.~A. Kochetov}{\jmp}{36}{4667}{1995}.

\bibitem{vvps95} \citn{V.~R. Vieira and P.~D. Sacramento}{\npb}{448}{331}{1999}.

\bibitem{spg00} M.~Stone, K.~S. Park, and A.~Garg, J. Math. Phys. {\bf 41},
8025 (2000).

\bibitem{pb78} P.~A. Braun, Teor.\ Mat.\ Fizika {\bf 37}, 355
(1978) [Sov.\ Phys.\ Theor.\ Math.\ Phys.\ {\bf 37}, 1070 (1978)].

\bibitem{agjmp00} A.~Garg, math-ph/0003005.

\bibitem{ch62} Conyers Herring, Rev.\ Mod.\ Phys.\ {\bf 34}, 631 (1962).

\bibitem{llqm2} See \rno{llqm}, Sec.~50, problem 3.

\bibitem{agajp00} \citn{A.~Garg}{Amer.\ J.\ Phys.\ }{68}{430}{2000}.

\end{references}
\end{document}